\documentclass[12pt,preprint]{aastex}

\slugcomment{In preparation for The Astrophysical Journal}%
\shorttitle{Fitting Narrow Emission Lines}%
\shortauthors{T.~Park et~al.}%

\begin{document}

\newcommand\obs{{\rm obs-id}}
\newcommand\up{\raisebox{1.7ex}[0pt]}
\newcommand\Case{C{\scriptsize ASE}}
\newcommand\Cases{C{\scriptsize ASES}}
\newcommand\model{M{\scriptsize ODEL}}
\renewcommand\models{M{\scriptsize ODELS}}
\newcommand\iid{\stackrel{\rm iid}{\sim}}
\newcommand\Estep{E-{\scriptsize STEP}}
\newcommand\Mstep{M-{\scriptsize STEP}}
\newcommand\step{S{\scriptsize TEP}}
\newcommand\Ymis{Y_{\rm mis}}
\newcommand\Yobs{Y_{\rm obs}}
\newcommand\Yaug{Y_{\rm aug}}

%%%%%%%%%%%%%%%%%%%%%%%%%%%%%%%%%%%%%%%%%%%%%%%%%%%%%%%%%%%%%%%%%%%%%%%%%%%%%%

\title{\bf Searching for Narrow Emission Lines in X-ray Spectra: \\
           Computation and Methods}

\author{Taeyoung~Park\altaffilmark{1},
David~A.~van~Dyk\altaffilmark{2},
Aneta~Siemiginowska\altaffilmark{3}}

\affil{$^1$Department of Statistics\\
University of Pittsburgh \\
4200 Fifth Avenue, Pittsburgh, PA 15120
\email{tpark@pitt.edu}}

\affil{$^2$Department of Statistics\\
University of California, \\
364 ICS Bldg One, Irvine, CA 92697-1250
\email{dvd@ics.uci.edu}}

\affil{$^3$Smithsonian Astrophysical Observatory, \\
60 Garden Street, Cambridge, MA 02138
\email{aneta@head.cfa.harvard.edu}}

\begin{abstract}

The detection and quantification of narrow emission lines in X-ray
spectra is a challenging statistical task. The Poisson nature of the
photon counts leads to local random fluctuations in the observed
spectrum that often results in excess emission in a narrow band of
energy resembling a weak narrow line. From a formal statistical
perspective, this leads to a (sometimes highly) multimodal
likelihood. Many standard statistical procedures are based on
(asymptotic) Gaussian approximations to the likelihood and simply
cannot be used in such settings. Bayesian methods offer a more
direct paradigm for accounting for such complicated likelihood
functions but even here multimodal likelihoods pose significant
computational challenges. The new Markov chain Monte Carlo (MCMC)
methods developed in \citet{vand:park:08} and \citet{park:vand:08},
however, are able to fully explore the complex posterior
distribution of the location of a narrow line, and thus provide
valid statistical inference. Even with these computational tools,
standard statistical quantities such as means and standard
deviations cannot adequately summarize inference and standard
testing procedures cannot be used to test for emission lines.  In
this paper, we use new efficient MCMC algorithms to fit the location
of narrow emission lines, we develop new statistical strategies for
summarizing highly multimodal distributions and quantifying valid
statistical inference, and we extend the method of posterior
predictive p-values proposed by \citet{prot:etal:02} to test for the
presence of narrow emission lines in X-ray spectra. We illustrate
and validate our methods using simulation studies and apply them to
the {\it Chandra} observations of the high redshift quasar
PG1634+706.

\end{abstract}


%========================================================================

\section{Introduction}
\label{park:sec:intro}

\subsection{Scientific Background}
\label{park:sec:sci}

Modern X-ray observations show complex structures in both the
spatial and spectral domains of various astrophysical sources.
Nonetheless, active galalactic nuclei (AGN) including quasars'
nuclei remain spatially unresolved even with the highest-resolution
X-ray telescopes. Most of their energy is released within the
unresolved core, and only spectral and timing information is
available to study the nature of the X-ray emission. Generally
speaking, emission and absorption lines constitute an important part
of the X-ray spectrum in that they can provide information as to the
state of plasma. One of the goals of X-ray data analysis is to
understand the components present in the spectrum, and to obtain
information about the emission and absorption features, as well as
their locations and relation to the primary quasar emission. The
detection of weak lines in noisy spectra is the main statistical
problem in such analyses: Is a bump observed in the spectrum related
to a real emission line or is it simply an artifact of the
Poissonian noise?

Although quasars' X-ray spectra are usually featureless as
expected based on the Comptonization process \citep[see for
example][]{mark:nowa:wilm:05,sobo:siem:zyck:04,siko:etal:97}, an
important X-ray emission feature identified in AGN and quasars
spectra is the iron K emission line \citep[see recent review
by][]{miller:07}. Determining the origin and the nature of this
line is one of main issues in AGN and quasar research.  This line
is thought to come directly from illuminated accretion flow as a
fluorescent process \citep{fabi:06}. The location of the line in
the spectrum indicates the ionization state of iron in the
emitting plasma, while the width of the line tells us the velocity
of the plasma \citep{fabi:06}. The iron line provides a direct
probe of the innermost regions of accretion flow and matter in
close vicinity of a black hole.

Absorption features associated with the outflowing matter (warm
wind, partial covering absorber) have also been observed in recent
X-ray observations \citep{gall:etal:02,char:etal:02,poun:reev:07}.
Although the location and width of absorption lines provide
information as to the velocity of the absorber and its distance
from the quasar, this article focuses on statistical issues in
fitting the spectral location of narrow emission lines, i.e.,
identifying the ionization state.

There are two parts to the Fe-K-alpha emission line observed in
AGN \citep{yaqoob:etal:01}: one is a broad component thought to be
a signature of a relativistic motion in the innermost regions of
an accretion flow; the other is a narrow component that is a
result of a reflection off the material at larger distances from
the central black hole.  A detection of the broad component is
challenging as it requires a spectral coverage over a large energy
range, so the continuum emission is well determined and the broad
line can be separated \citep{reeves:etal:06}. The relativistic
line profile is broad and skewed, and two strong peaks of the
emission line that originates in a relativistic disk can be
prominent and narrow. While the full profile of the broad line may
not be easily separable from the continuum, these two peaks may
provide a signature for this line in the X-ray spectrum.  The
broad Fe-line gives an important diagnostic of the gas motion and
can be used to determine the spin of a black hole
\citep{miller:06}; see also an alternative model for the ``red
wing" component by \citet{miller:etal:08}. The narrow component of
the line gives diagnostics of the matter outside the accretion
disk and conditions at larger distances from the black hole; see
Fe-line Baldwin effect discussion in \citet{jiang:etal:06}. Both
line components are variable and the line may ``disappear'' from
the spectrum \citep{yaq:etal:01}.

The spectral resolution of X-ray CCD detectors (for example
100-200~eV in ACIS on {\it Chandra} or EPIC on XMM-{\it Newton})
is relatively low with respect to the predicted width of narrow
($< 5000-30,000$~km~s$^{-1}$) emission or absorption lines in AGN
and quasars. Observations with grating instruments (RGS or HEG)
can provide high resolution X-ray spectra, but the effective area
of the present X-ray telescopes is too low for efficient AGN
detections, and only a handful of bright low redshift sources have
been observed with gratings to date \citep{yaqoob:07}. Therefore
mainly the X-ray CCD spectra of lower resolution are used to study
large samples of AGN and quasars \citep[see for
example][]{guai:06,page:etal:05,jimen:etal:05}.

Using these relatively low resolution X-ray detectors, the
Fe-K-alpha emission line can be narrow enough to be contained
entirely in a single detector bin.  In some cases (for example in
{\it Chandra}) the line may occupy a few bins.  In this article we
focus on the statistical problem of fitting the spectral location
of an emission line or a set of emission lines that are narrow.
This is a common objective in high-energy analyses, but as we
shall discuss fitting these relatively narrow features poses
significant statistical challenges. In particular we find evidence
that using line profiles that are narrower than we actually expect
the emission line to be can improve the statistical properties of
the fitted emission line location.

%======================================================================

\subsection{X-Ray Spectral Analysis}
\label{park:sec:hea}

X-ray spectra, such as those available with the {\it Chandra X-ray
Observatory} carry much information as to the quasar's physics.
Taking advantage of the spectral capacity of such instruments,
however, requires careful statistical analysis. For example, the
resolution of such instruments corresponds to a fine discretization
of the energy spectrum. As a result, we expect a low number of
counts in each bin of the X-ray spectrum.  Such low-count data make
the Gaussian assumptions that are inherent in traditional minimum
$\chi^2$ fitting inappropriate. A better strategy, which we employ,
explicitly models photon arrivals as an inhomogeneous Poisson
process \citep{vand:etal:01}.  In addition, data are subject to a
number of processes that significantly degrade the source counts,
e.g., the absorption, non-constant effective area, blurring of
photons' energy, background contamination, and photon pile-up. Thus,
we employ statistical models that directly account for these aspects
of data collection. In particular, we design a highly structured
multilevel spectral model with components for both the data
collection processes and the complex spectral structures of the
sources themselves. In this highly structured spectral model, a
Bayesian perspective renders straightforward methods that can handle
the complexity of {\it Chandra} data
\citep{vand:etal:01,vand:kang:04,vand:etal:06}. As we shall
illustrate, these methods allow us to use low-count data, to search
for the location of a narrow spectral line, to investigate its
location's uncertainty, and to construct statistical tests that
measure the evidence in the data for including the spectral line in
the source model.

%======================================================================

\subsection{A Statistical Model for the Spectrum}
\label{park:sec:model}

The energy spectrum can be separated into two basic parts: a set of
continuum terms and a set of several emission lines\footnote{The
model can be extended to account for absorption lines, but in this
paper we focus on additive features such as emission lines.}. We
begin with a standard spectral model that accounts for a single
continuum term along with several spectral lines. Throughout this
paper, we use $\theta$ as a general representation of model
parameters in the spectral model. The components of
$\theta=(\theta^C,\theta^L,\theta^A,\theta^B)$ represent the
collection of parameters for the Continuum, (emission) Lines,
Absorption, and Background contamination, respectively. (Notice that
the roman letters in the superscripts serve as a mnemonic for these
four processes.) Because the X-ray emission is measured by counting
the arriving photons, we model the expected Poisson counts in energy
bin $j\in{\cal J}$, where ${\cal J}$ is the set of energy bins, as
\begin{eqnarray}
  \Lambda_j(\theta) \ =\ \Delta_j f\big(\theta^C,E_j\big)+
    \sum\limits_{k=1}^K\lambda_k\pi_j\big(\mu_k,\nu_k\big),
  \label{park:eq:ideal}
\end{eqnarray}
where $\Delta_j$ and $E_j$ are the width and mean energy of bin
$j$, $f(\theta^C,E_j)$ is the expected counts per unit energy due
to the continuum term at energy $E_j$, $\theta^C$ is the set of
free parameters in the continuum model, $K$ is the number of
emission lines, $\lambda_k$ is the expected counts due to the
emission line $k$, and $\pi_j(\mu_k,\nu_k)$ is the proportion of
an emission line centered at energy $\mu_k$ and with width $\nu_k$
that falls into bin $j$. There are a number of smooth parametric
forms to describe the continuum in some bounded energy range; in
this article we parameterize the continuum term $f$ as a power
law, i.e., $f(\theta^C,E_j)=\alpha^CE_j^{-\beta^C}$ where
$\alpha^C$ and $\beta^C$ represent the normalization and photon
index, respectively. The emission lines can be modeled via the
proportions $\pi_j(\mu_k,\nu_k)$ using narrow Gaussian
distributions, Lorentzian distributions, or delta functions; the
counts due to the emission line are distributed among the bins
according to these proportions. While the Gaussian or Lorentzian
function parameterizes an emission line in terms of center and
width, the center is the only free parameter with a delta
function; the width of the delta function is effectively the width
of the energy bin in which it resides.

While the model in Equation~\ref{park:eq:ideal} is of primary
scientific interest, a more complex statistical model is needed to
address the data collection processes mentioned in
\S\ref{park:sec:hea}. We use the term {\it statistical model} to
refer to the model that combines the {\it source} or {\it
astrophysical model} with a model for the stochastic processes
involved in data collection and recording. Thus, in addition to the
source model, the statistical model describes such processes as
instrument response and background contamination.  Specifically, to
account for the data collection processes,
Equation~\ref{park:eq:ideal} is modified via
\begin{eqnarray}
  \Xi_l(\theta) \ =\
    \sum\limits_{j\in{\cal J}} M_{lj}\Lambda_j(\theta)d_j u(\theta^A,E_j)+\theta_l^B
  \label{park:eq:obs}
\end{eqnarray}
where $\Xi_l(\theta)$ is the expected observed Poisson counts in
detector channel $l\in{\cal L}$, ${\cal L}$ is the set of detector
channels, $M_{lj}$ is the probability that a photon that arrives
with energy corresponding to bin $j$ is recorded in detector
channel $l$ (i.e., $\mathbf{M}=\{M_{lj}\}$ is the so-called
redistribution matrix or RMF commonly used in X-ray analysis),
$d_j$ is the effective area (i.e., ARF, a calibration file
associated with the X-ray observation) of bin $j$,
$u(\theta^A,E_j)$ is the probability that a photon with energy
$E_j$ is {\it not} absorbed, $\theta^A$ is the collection of
parameters for absorption, and $\theta_l^B$ is a Poisson intensity
of the background counts in channel $l$. While the scatter
probability $M_{lj}$ and the effective area $d_j$ are presumed
known from calibration, the absorption probability is
parameterized using a smooth function; see \citet{vand:hans:02}
for details. To quantify background contamination, a second data
set is collected that is assumed to consist only of background
counts; the background photon arrivals are also modeled as an
inhomogeneous Poisson process.

%======================================================================

\subsection{Difficulty with Identifying Narrow Emission Lines}
\label{park:sec:narrow}

Unfortunately, the statistical methods and algorithms developed in
\citet{vand:etal:01} cannot be directly applied to fitting {\it
narrow} emission lines. There are three obstacles that must be
overcome in order to extend Bayesian highly structured models to
spectra containing narrow lines. In particular, we must develop (1)
new computational algorithms, (2) statistical summaries and methods
for inference under highly multimodal posterior distributions, and
(3) statistical tests that allow us to quantify the statistical
support in the data for including an emission line or lines in the
model. Our main objective in this paper is to extend the methods of
\citet{vand:etal:01} in these three directions, and to evaluate and
illustrate our proposals. Here we discuss each of these challenges
in detail.

\paragraph{Challenge 1: Statistical Computation.}
Fitting the location of narrow lines requires new and more
sophisticated computational techniques than those developed by
\citet{vand:etal:01}. Indeed, the algorithms that we develop require
a new theoretical framework for statistical computation: they are
not examples of any existing algorithm with known properties.
Although the details of this generalization are well beyond the
scope of this article, we can offer a heuristic description; a more
detailed description is given in Appendix~\ref{ap:alg}. Readers who
are interested in the necessary theoretical development of the
statistical computation techniques are directed to
\citet{vand:park:04,vand:park:08} and \citet{park:vand:08}.

The algorithms used by \citet{vand:etal:01} to fit the structured
Bayesian model described in \S\ref{park:sec:model} are based on
the probabilistic properties of the statistical models. For
example, the parameters of a Gaussian line profile can be fit by
iteratively attributing a subset of the observed photons to the
line profile and using the mean and variance of these photon
energies to update the center and width of the line profile.  The
updated parameters of the line profile are used to again attribute
a subset of the photons to the line, i.e., to stochastically
select a subset of the photons that are likely to have arisen out
of the physical processes at the source corresponding to the
emission line.  These algorithms are typically very stable. For
example, they only return statistically meaningful parameters
because the algorithms themselves mimic the probabilistic
characteristics of the statistical model. The family of
Expectation/Maximization (EM) algorithms \citep{demp:lair:rubi:77}
and Markov chain Monte Carlo (MCMC) methods such as the Gibbs
sampler \citep{gema:gema:84} are the examples of statistical
algorithms of this sort.

A drawback of these algorithms is that in some situations they can be
slow to converge. When fitting the location of a Gaussian emission
line, for example, the location is updated more slowly if the line
profile is narrower. This is because only photons with energies very
close to the current value of the line location can be attributed to
the line. Updating the line location with the mean of the energies of
these photons cannot result in a large change in the emission line
location. The situation becomes chronic when a delta function is used
to model the line profile: The line location parameter sticks at its
starting value throughout the iteration.

It is to circumvent this difficulty that we develop both new EM-type
algorithms \citep{vand:park:04} and new MCMC samplers specially
tailored for fitting narrow lines.  Our new samplers are motivated
by the Gibbs sampler, but constitute a non-trivial generalization of
Gibbs sampling known as partially collapsed Gibbs sampling
\citep{vand:park:08,park:vand:08}; see Appendix~\ref{ap:alg}. Our
updated versions of both classes of algorithms are able to fit
narrow lines by avoiding the attribution of photons to the emission
line during the iteration. Such algorithms tend to require fewer
iterations to converge regardless of the width of the emission line.
Because they involve additional evaluation of quantities evolving
the large dimensional redistribution matrix, $M$, however, each
iteration of these algorithms can be significantly more costly in
terms of computing time. A full investigation of the relative merit
of the algorithms and a description of how the computational
trade-offs can be played to derive optimal algorithms are beyond the
scope of this paper. Except in Appendix~\ref{ap:alg}, we do not
discuss the details of the algorithms further in this article;
interested readers are directed to \citet{vand:park:04,vand:park:08}
and \citet{park:vand:08}.

\paragraph{Challenge 2: Multimodal Likelihoods.}
The likelihood function for the emission line location(s) is highly
multimodal.  Each mode corresponds to a different relatively likely
location for an emission line or a set of emission lines. Standard
statistical techniques such as computing the estimates of the line
locations with their associated error bars or confidence intervals
implicitly assume that the likelihood function is unimodal and bell
shaped. Because this assumption is clearly and dramatically
violated, these standard summary statistics are unreliable and
inadequate.

Unfortunately, there are no readily available and generally
applicable simple statistical summaries to handle highly
multimodal likelihoods. Instead we must develop summaries that are
tailored to the specific scientific goals in a given analysis.
Because general strategies for dealing with multimodal likelihood
functions are little known to astronomers and specific strategies
for dealing with multimodal likelihood functions for the location
of narrow spectral lines do not exist, one of the primary goals of
this article is to develop and illustrate these methods.

A fully Bayesian analysis of our spectral model with narrow
emission lines is  computationally demanding, even with our new
algorithms. Thus, we develop techniques that are much quicker and
give similar results for the location of emission lines. These
methods based on the so-called {\it profile posterior
distribution} do not stand on as firm of a theoretical footing as
a fully Bayesian analysis, but are much quicker and thus better
suited for {\it exploratory data analysis}. The profile posterior
distribution along with our exploratory methods are fully
described and compared with the more sophisticated Bayesian
analysis.

\paragraph{Challenge 3: Testing for the Presence of Narrow Lines.}

In addition to fitting the location of one or more emission lines, we
often would like to perform a formal test for the inclusion of the
emission lines in the statistical model. That is, we would like to
quantify the evidence in a potentially sparse data set for a
particular emission line in the source.

Testing for a spectral line is an example of a notoriously
difficult statistical problem in which the standard theory does
not apply. There are two basic technical problems. First, the
simpler model that does not include a particular emission line is
on the boundary of the larger model that does include the line.
That is, the intensity parameter of an emission line is zero under
the simpler model and cannot be negative under the larger model.
An even more fundamental problem occurs if either the line
location or width is fit, because these parameters have {\it no
value} under the simpler model. The behavior (i.e., sampling
distribution) of the likelihood ratio test statistic under the
simpler model is not well understood and cannot be assumed to
follow the standard $\chi^2$ distribution, even asymptotically.
\citet{prot:etal:02} propose a Monte-Carlo-based solution to this
problem based on the method of posterior predictive p-values
\citep{rubi:84,meng:94a}.  In this article we extend the
application of Protassov~{\it et~al}.'s solution to the case when
we fit the location of a narrow emission line, a situation that
was avoided in \citet{prot:etal:02}.

\subsection{Outline of the Article}
\label{park:sec:outline}

The remainder of the article is organized into four sections.
\S\ref{park:sec:model-based} reviews Bayesian inference and Monte
Carlo methods with an emphasis on  multimodal distributions,
outlines our computation methods, proposes new summaries of
multimodal distributions, and describes exploratory statistical
methods in this setting. We introduce illustrative examples in
\S\ref{park:sec:model-based}, but detailed spectral analysis is
postponed in order to allow us to focus on our proposed methods. In
\S\ref{park:sec:simul}, a simulation study is performed to
investigate the statistical properties of our proposed methods, with
some emphasis placed on the potential benefits of model
misspecification. \S\ref{park:sec:quasar} presents the analysis of
the high redshift quasar PG1634+706, and how to test for the
inclusion of the line in the spectral model. Concluding remarks
appear in \S\ref{park:sec:conclusion}. An appendix outlines the
computational methods we developed specifically for fitting the
location of narrow emission lines.

%======================================================================

\section{Model-Based Statistical Methods}
\label{park:sec:model-based}

\subsection{Likelihood-Based and Bayesian Methods}
\label{park:sec:bayes}

Using a Poisson model for the photon counts, the likelihood function
of the parameter in the spectral model described in
\S\ref{park:sec:model} is given by $L(\theta|\mathbf{Y}^{\rm
obs})\propto\prod_{l\in{\cal L}}\Xi_l(\theta)^{{Y}^{\rm obs}_l}
\exp[-\Xi_l(\theta)]$ where $\mathbf{Y}^{\rm obs}=\{{Y}^{\rm
obs}_l,l\in{\cal L}\}$ denotes the observed photon counts. With
likelihood-based methods, the parameter value that maximizes the
probability of the observed data is generally chosen as an estimate
of $\theta$; this estimate is called the maximum likelihood estimate
(MLE). In Bayesian methods, prior knowledge for $\theta$ can be
combined with the information in the observed data. A {\it prior
distribution} can be used to quantify information from other sources
or to impose structure on a set of parameters. The prior
distribution is combined with the likelihood to form a {\it
posterior distribution}.  The prior distribution is denoted by
$p(\theta)$ and the posterior distribution by
$p(\theta|\mathbf{Y}^{\rm obs})$. Bayesian inferences for $\theta$
are based on the posterior distribution. Using Bayes' theorem, the
prior distribution and the likelihood function are combined to form
the posterior distribution via
\begin{eqnarray}
  p(\theta|\mathbf{Y}^{\rm
obs})
  &=& \frac{p(\theta)L(\theta|\mathbf{Y}^{\rm
obs})}{\int p(\theta)L(\theta|\mathbf{Y}^{\rm
obs})d\theta}\nonumber\\
  &\propto& p(\theta)L(\theta|\mathbf{Y}^{\rm
obs}),
  \label{park:eq:Bayes}
\end{eqnarray}
where the last proportionality holds because $p(\mathbf{Y}^{\rm
obs})=\int p(\theta)L(\theta|\mathbf{Y}^{\rm obs})d\theta$ does
not depend on $\theta$ and, given the observed data, is considered
a constant.

\begin{figure}[t]
\begin{center}
  \includegraphics[width=6in]{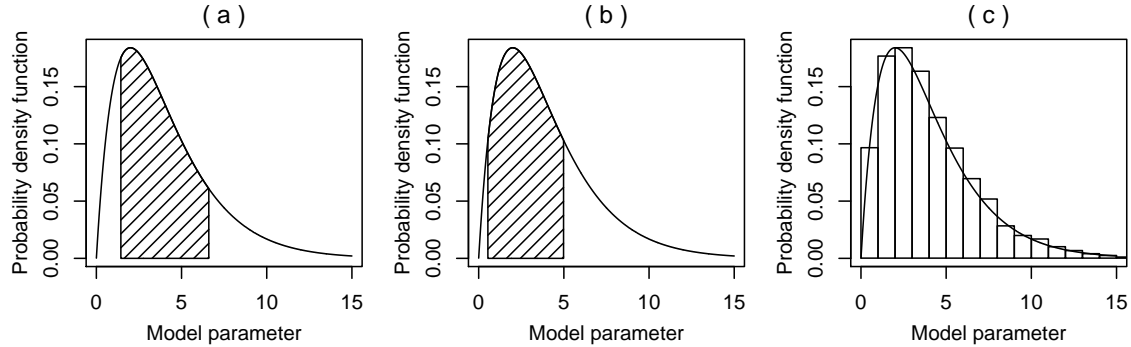}
    \caption{Various Summaries of a Right-skewed Distribution.
    Panels~(a) and~(b) illustrate the 68\% equal-tail interval and
    the 68\% HPD interval, respectively. Panel~(c) shows that a
    theoretical probability density function agrees with
    its Monte Carlo simulation.}
    \label{park:fig:skewed}
\end{center}
\end{figure}

Bayesian statistical inferences are made in terms of probability
statements, which are quantified using various numerical summaries
of the posterior distribution. To illustrate this, we consider a
stylized right-skewed distribution; see
Figure~\ref{park:fig:skewed}. This distribution is similar to the
posterior distribution of the expected counts due to the emission
line~$k$, $\lambda_k$, because this parameter is necessarily
non-negative.

Although from a Bayesian perspective the posterior distribution of a
model parameter is the complete summary of statistical inference for
that parameter, it is often useful to summarize the posterior
distribution using point estimates or intervals. Commonly used
Bayesian point estimates of a parameter are the mean, median, and
mode(s) of the posterior distribution. Error bars for the point
estimates can be computed based on the variation of the posterior
distribution. The equal-tail interval and the Highest Posterior
Density (HPD) interval are both commonly used summaries of
uncertainty. For example, a 68\% equal-tail posterior interval is
the central interval of the posterior distribution and corresponds
to the range of values of the parameter above and below which lies
exactly 16\% of the posterior probability. A 68\% HPD interval, on
the other hand, is the interval of values that contains 68\% of
posterior probability and within which the density is never lower
than that outside the interval. The 68\% HPD interval is the
shortest possible interval that accounts for 68\% of the posterior
probability. This is illustrated in Figure~\ref{park:fig:skewed}(a)
and~(b). The equal-tail interval achieves the same probability as
the HPD interval by excluding a more likely region and by including
a less likely region. When the posterior distribution is unimodal
and symmetric, the equal-tail interval and the HPD interval are
identical.

In addition to computing parameter estimates and their error bars,
it is often important to check if model assumptions are supported
by the data. One way to do this is to generate simulated data
under the model and compare the simulated data with the observed
data; refer to \citet{prot:etal:02} where this strategy is used to
determine whether emission line profiles should be included in the
spectral model for a gamma-ray burst. If the simulated data vary
systematically for the observed data, it is an indication that the
model used to simulate the data may not be adequate. In a Bayesian
analysis, we might generate such simulated data from the {\it
posterior predictive distribution}, denoted by
$p(\widetilde{\mathbf{Y}}|\mathbf{Y}^{\rm obs})$, i.e.,
\begin{eqnarray}
  p(\widetilde{\mathbf{Y}}|\mathbf{Y}^{\rm obs})
    &=& \int p(\widetilde{\mathbf{Y}}|\mathbf{Y}^{\rm obs},\theta)p(\theta|\mathbf{Y}^{\rm obs})d\theta \nonumber\\
    &=& \int L(\theta|\widetilde{\mathbf{Y}})p(\theta|\mathbf{Y}^{\rm obs})d\theta,
  \label{park:eq:post-pred}
\end{eqnarray}
where $\widetilde{\mathbf{Y}}$ represents an unknown future
observation and the last equation follows because
$\widetilde{\mathbf{Y}}$ and $\mathbf{Y}^{\rm obs}$ are
conditionally independent given $\theta$. In words, the posterior
predictive distribution averages the likelihood function over the
posterior distribution of $\theta$. Data are simulated from the
posterior predictive distribution and then used to make predictive
inferences; see \S\ref{park:sec:model-check} for details.

Posterior simulation plays a central part in applied Bayesian
analysis because of the usefulness of a simulation that can often be
relatively easily generated from a posterior distribution. In
performing simulations, given a large enough sample, a histogram of
a Monte Carlo simulation can provide practically complete
information about an actual posterior distribution.
Figure~\ref{park:fig:skewed}(c)
shows that the histogram of the Monte Carlo simulation carries the
same information as the posterior distribution itself. Thus, once a
Monte Carlo simulation is obtained, it can be used to compute the
mean, variance, percentiles, and other summaries of the posterior
distribution. In particular, with a random simulation of size $N$,
the posterior mean can be approximated as
\begin{eqnarray}
  {\rm E}(\theta|\mathbf{Y}^{\rm obs}) \ =\ \int \theta p(\theta|\mathbf{Y}^{\rm obs})d\theta
  \ \approx\ \frac{1}{N}\sum_{\ell=1}^N \theta^{(\ell)},
  \label{park:eq:mean}
\end{eqnarray}
where $\{\theta^{(\ell)}, \ell=1,\dots,N\}$ is a simulation from
the posterior distribution, $p(\theta|\mathbf{Y}^{\rm obs})$.
\begin{figure}[p]
\begin{center}
  \includegraphics[width=5in]{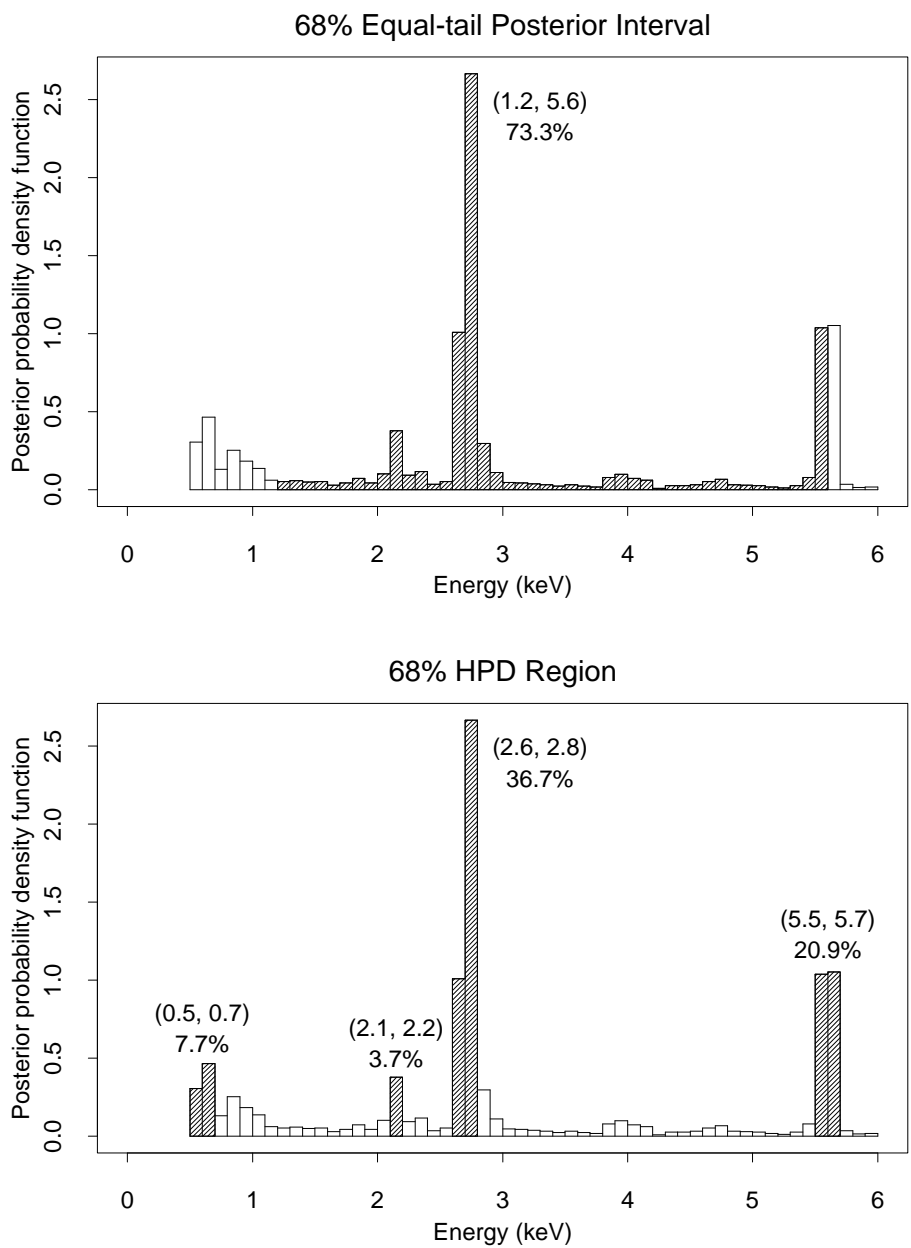}
    \caption{Comparison of an Equal-tail Interval and an HPD Region
    Computed with a Monte Carlo Simulation. The shaded interval in
    the top panel indicates a 68\% equal-tail interval and the shaded
    region in the bottom panel indicates a 68\% HPD region. On the
    top of each shaded area, its range of energies and the
    corresponding posterior probability accounted for are shown. The
    HPD region is not an interval, and is much more informative as to
    the likely values of the line location. Here, we use
    a coarse binning for illustrative purposes. In our actual analysis, we
    use finer binning to construct more precise equal-tail intervals
    and HPD regions. The shape of the histogram is typical of what
    we observe for the location of a relatively weak emission line.}
    \label{park:fig:HPD}
\end{center}
\end{figure}
A 68\% equal-tail posterior interval is computed by generating a
Monte Carlo simulation of size $N$ from the posterior
distribution, sorting the simulated values into increasing order,
and choosing the $[0.16N]$th and the $[0.84N]$th values in the
list. With the Monte Carlo simulation, a 68\% HPD region is
computed by segmenting the range of possible parameter values into
bins, approximating the posterior probability of each bin as the
proportion of the simulated values in that bin, and computing a
region by beginning with the bin with the largest posterior
probability and adding additional bins in the order of their
posterior probabilities until the resulting region contains at
least 68\% of the posterior probability.

%============================================

\subsection{Outline of Computational Strategies}
\label{park:sec:outline-method}

Our search for a narrow emission line begins by finding modes of its
posterior distribution that correspond to plausible locations of an
emission line. To find the modes (and to compte the profile
posterior distribution), we use an algorithm optimized for this
problem (i.e., the Rotation(9) EM-type algorithm, see
Appendix~\ref{ap:alg} and \citet{vand:park:04} for details). Because
the posterior distribution of the line location is highly multimodal
(see \S\ref{park:sec:mm}), the algorithm is run using multiple
starting values selected across the entire energy range of possible
line locations, e.g., 50 starting values for the line location
equally spaced between 1.0~keV and 6.0~keV. Using multiple starting
values enables us to identify the important local modes of the
posterior distribution. It is important to use enough starting
values to ensure all of the important modes are identified. This is
a standard strategy, long advocated in texts on Bayesian data
analysis \citep{gelm:etal:95} and is closely related to the
computation of the profile posterior distribution described in
\S\ref{park:sec:advise}. The profile posterior distribution is
computed by fixing the line location at each value of a fine grid,
finding the posterior modes of the other model parameters, and
plotting the resulting maximum posterior probability as a function
of the line location. This procedure corresponds to the projected
delta-chi-square method in the chi-square setting, see
\citet{lamp:etal:76} and \citet{pres:etal:92}. Mode finding also
begins with a fine grid of starting values, but we run the mode
finder allowing all parameters including the line location to be
fit.

After the modes are found, Monte Carlo simulation techniques
optimized to this problem can be run to further investigate the
uncertainty of the possible line locations. We employ
state-of-the-art MCMC samplers, i.e., Partially Collapsed Gibbs
(PCG) I for the delta function emission line and PCG II for the
Gaussian emission line to obtain the posterior distribution of line
location; see Appendix~\ref{ap:alg}, \citet{vand:park:08}, and
\citet{park:vand:08} for details. To ensure the convergence of a
Markov Chain constructed by the MCMC samplers, we run multiple
chains with overdispersed starting values (e.g., 1~keV, 3~keV, and
5~keV for the line location parameter) and monitor the convergence
qualitatively and quantitatively. For example, we compute the
estimate of the potential scale reduction \citep{gelm:rubi:92},
denoted by ${\widehat{R}^{1/2}}$, for all parameters of interest. If
${\widehat{R}^{1/2}}$ is near 1 (e.g., below 1.2) for each of the
parameters, we collect the second halves of the chains together and
use these Monte Carlo draws for inference; see \citet{gelm:rubi:92}
for theoretical justification and discussion. There are of course
many strategies that one might employ to construct efficient Monte
Carlo samplers. Methods based on annealing or tempering, or that use
explicitly parallel methods are often useful for exploring
multimodal posterior distributions. Given the low autocorrelation of
the simulated values produced by our method, we have not pursued
these strategies.

%============================================

\subsection{Summarizing Multimodal Posterior Distributions}
\label{park:sec:mm}

Fitting a narrow emission line often tends to yield a highly
multimodal posterior distribution of the line location, as shown
in Figure~\ref{park:fig:HPD}. Thus, we are interested in the two
types of intervals for the posterior distributions: an equal-tail
posterior interval and an HPD region. Figure~\ref{park:fig:HPD}
illustrates a 68\% equal-tail posterior interval and a 68\% HPD
region for the line location when its posterior distribution is
highly multimodal. This is only an example of a highly multimodal
posterior distribution. We come to the details of our analysis of
a line location in \S\ref{park:sec:simul}. The 68\% equal-tail
posterior interval is a central interval, so that it includes
segments with nearly zero posterior probability here and there
within the interval. Because the posterior distribution of the
line location in Figure~\ref{park:fig:HPD} is multimodal, the 68\%
HPD region consists of the four shaded disjoint intervals; notice
that the height of each of the histogram bars outside the HPD
region is less than that of those within the region. In such a
multimodal posterior distribution, the HPD region not only is
shorter in length but also conveys more information as to likely
locations of the line than does the equal-tail interval.

\begin{figure}[p]
\begin{center}
  \includegraphics[width=5in]{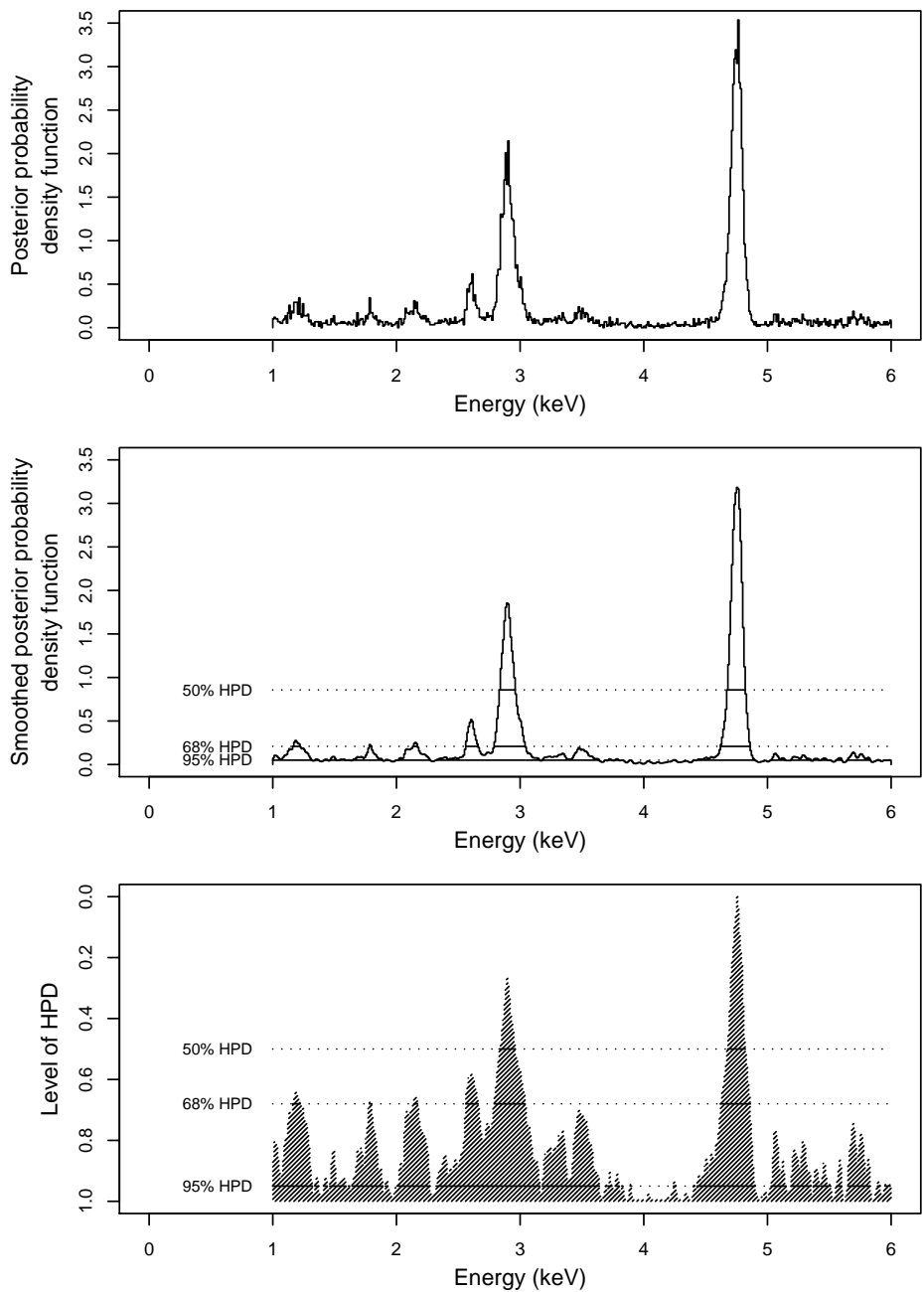}
    \caption{Comparing an Unsmoothed Marginal Posterior Distribution
    with its HPD Graph computed with Gaussian Kernel Smoothing.  The
    top panel presents an unsmoothed posterior distribution computed
    from Monte Carlo draws, plotted with the same resolution as the
    {\it Chandra} energy spectrum.  In the middle panel, the posterior
    distribution is smoothed using Gaussian kernel smoothing with
    standard deviation 0.01~keV, the width of one energy bin in the
    {\it Chandra} spectrum.  Based
    on the smoothed posterior distribution, we compute 100 HPD regions
    with levels ranging from 1\% to 100\%. These intervals are plotted
    against their levels in the HPD graph shown in the bottom panel.  The
    horizontal solid lines in the middle and bottom panels are the 50\%,
    68\%, and 95\% HPD regions, and the horizontal dotted lines are
    the intervals outside each HPD region.}
    \label{park:fig:HPD-illust}
\end{center}
\end{figure}

In Figure~\ref{park:fig:HPD}, we use a coarse binning to illustrate
the distinction between the equal-tail posterior interval and HPD
region. This results in a posterior region that is relatively
imprecise; in our spectral analysis, we use a finer binning to
construct a more precise region from a Monte Carlo simulation. When
the posterior distribution of the line location is plotted with the
same fine binning as the {\it Chandra} energy spectrum, however, it
may not be smooth due to Monte Carlo errors, as shown in the top panel
of Figure~\ref{park:fig:HPD-illust}.  An HPD region computed
from the unsmoothed posterior distribution may result in a combination
of too many posterior intervals. To avoid such fragmentation of HPD
regions, we use Gaussian kernel smoothing to smooth the posterior
distribution before we compute HPD regions. The middle panel of
Figure~\ref{park:fig:HPD-illust} presents the smoothed posterior
distribution resulting from applying Gaussian kernel smoothing with
standard deviation equal to the bin size of the {\it Chandra} energy
spectrum, i.e., 0.01~keV. The smoothed posterior distribution smooths
out lower posterior probabilities but does not flatten higher
posterior probabilities too much.

We propose a new graphical summary to better describe the HPD
regions of a (smoothed) multimodal posterior distribution. An {\it
HPD graph} is constructed by plotting a series of HPD regions
against their corresponding HPD levels. For example, for the data
set used to compute the posterior distribution in
Figure~\ref{park:fig:HPD-illust}, we compute 100 HPD regions, one
for each of 100 levels, 1\%, 2\%, $\ldots$, and 100\%. Each of these
regions is a union of possible values of the line location. We can
plot the line location on the horizontal axis and the level on the
vertical axis. Each of the 100 HPD regions can then be plotted as a
union of horizonal line intervals at the appropriate level of the
HPD region on the vertical axis. The resulting HPD graph lets us
visualize many HPD regions with varying levels, so that all the
important modes of a multimodal distribution can be effectively
summarized with their relative posterior probabilities.  As an
illustration, we computed the three HPD regions with levels 50\%,
68\%, and 95\%, and plotted them in the middle panel of
Figure~\ref{park:fig:HPD-illust} along with the smoothed posterior
distribution.  The solid lines indicate the disjoint intervals that
compose each HPD region, and the dotted lines the intervals outside
the HPD region. The bottom panel in Figure~\ref{park:fig:HPD-illust}
shows the HPD graph in grey with the three HPD intervals from the
middle plot superimposed.

\begin{figure}[p]
\begin{center}
  \includegraphics[width=6in]{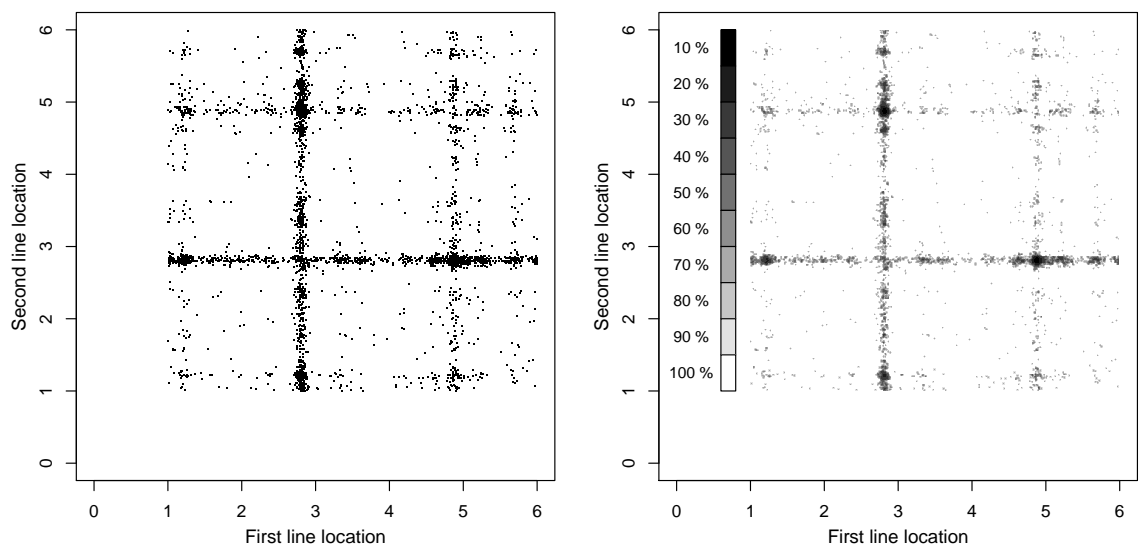}
    \caption{Comparing an Unsmoothed Joint Posterior Distribution
    with its 2-D HPD Graph computed with Bivariate Gaussian
    Kernel Smoothing. The left panel presents an unsmoothed joint
    posterior distribution computed from Monte Carlo draws,
    plotted with the same resolution as the
    {\it Chandra} energy spectrum.  The unsmoothed posterior
    distribution is smoothed using bivariate Gaussian kernel smoothing
    with covariance matrix ${0.01^2~0~~\choose~0~~0.01^2}$, where
    each marginal standard deviation corresponds to the size of an energy
    bin in the {\it Chandra} spectrum.
    Based on the smoothed posterior distribution, we compute
    ten 2-D HPD regions with levels ranging from 10\% to 100\%.
    These regions are plotted with different shades of grey
    in the 2-D HPD graph shown in the right panel.}
    \label{park:fig:TwoD-HPD-illust}
\end{center}
\end{figure}

When there are two parameters of interest, the 1-D HPD graph can be
extended to a 2-D HPD graph. That is, a joint posterior distribution
is computed from a Monte Carlo simulation by using bivariate
Gaussian kernel smoothing and used to construct 2-D HPD regions with
various levels. These HPD regions can then be plotted with different
shades of grey. For example, Figure~\ref{park:fig:TwoD-HPD-illust}
shows the 2-D HPD graph computed for the joint posterior
distribution of two possible line locations; see
\S\ref{park:sec:simul2} for further discussion on the scatter plot.
The left panel of Figure~\ref{park:fig:TwoD-HPD-illust} is an
unsmoothed joint posterior distribution. After applying bivariate
Gaussian kernel smoothing, 10 HPD regions (one for each of 10
levels, 10\%, 20\%, ..., and 100\%) are computed and plotted with
different shades of grey from black (10\%) to white (100\%), as
shown in the right panel of Figure~\ref{park:fig:TwoD-HPD-illust}.
HPD regions with lower levels contain pixels with higher posterior
probabilities, so that darker pixels indicate more probable regions
of the two line locations.

Multimodal likelihoods and posterior distributions pose another
challenge for statistical analysis. The calibration of many
standard statistical procedures is based on the asymptotic
Gaussian nature of the likelihood and posterior distribution.
Thus, standard methods for computing error bars and confidence
intervals rely on the likelihood being at least approximately
Gaussian. When the likelihood function exhibits the multimodal
features that we see in Figures~\ref{park:fig:HPD-illust} and
\ref{park:fig:TwoD-HPD-illust}, these standard results simply do
not apply. One consequence of this is that the nominal level of an
interval may not match its frequency coverage. That is, if we were
able to repeat our observation many times, we would find that
percentage of intervals that contain the true line location might
differ from the nominal level of the interval. In
\S\ref{park:sec:simul}, we find that the posterior distribution is
highly multimodal when the true emission line is weak or there are
several emission lines in a spectrum. In this case, we also find
that by {\it misspecifying} the model we are able to improve the
frequency properties of our proposed method. In particular, we
find that using a delta function line profile improves the
properties of our procedure even when the true line has
appreciable width.

%%%%%%%%%%%%%%%%%%%%%%%%%%%%%%%%%%%%%%%%%%%%%%%%%%%%%%%%%%%%%%%%%%%%%%%%%%%%%%

\subsection{Exploratory Data Analysis}
\label{park:sec:advise}

As discussed in \S\ref{park:sec:mm}, the posterior distribution of a
narrow emission line location tends to be highly multimodal. In this
case, the profile posterior distribution can be used as a handy and
quick-to-compute summary of the posterior distribution to explore
the possible locations of a spectral line. The profile posterior
distribution is the posterior distribution of a parameter evaluated
at the values of the other parameters that maximize the posterior
distribution. The profile posterior distribution is a Bayesian
analogue to the profile likelihood, a standard likelihood method for
dealing with nuisance parameters; see \citet{venz:mool:88} or
\citet{crit:etal:88} for applications of the profile likelihood. In
the context of parameter estimation, the distribution of the minimum
$\chi^2$ statistic described by \citet{lamp:etal:76} is closely
related to the profile posterior distribution.

Generally we do not advocate using the profile posterior
distribution as a substitute for the marginal posterior distribution
because interval or region estimates computed with the profile
posterior distribution have rather unpredictable statistical
properties. The marginal posterior distribution is obtained by
integrating out (i.e., averaging over) the other parameters and
computing the marginal posterior distribution requires sophisticated
numerical integration methods such as MCMC especially when the
dimension of the nuisance parameters is large. However, the profile
posterior distribution can be computed without MCMC and, as an
analogue to the marginal posterior distribution, can be used to
roughly examine the posterior distribution of a model parameter.
Thus, we believe the profile posterior distribution is well suited
for the initial exploration of the data because it gives a clear and
reliable set of potential locations for spectral lines.

The profile posterior distribution can be computed on a fine grid of
the possible values of the line location by running an optimizer to
maximize over the other model parameters for each value of the line
location on the grid; we recommend using stable optimizers such as
EM-type algorithms \citep{vand:park:04}. The profile posterior
distribution of the line location computed in this way is
computationally less demanding and cheaper in terms of CPU time than
the marginal posterior distribution produced by Monte Carlo methods
such as MCMC.
%======================================================================

\section{Simulation Study}
\label{park:sec:simul}

Our simulation study is conducted to assess the validity of the
highly structured multilevel spectral model discussed in
\S\ref{park:sec:model}, to illustrate the connection between the
multimodal posterior distribution of a single narrow emission line
and evidence for multiple lines, to illustrate the possible
advantage of the misspecification of an emission line width, and
to illustrate the relationship among Gaussian line parameters.

We consider the following six cases that we believe are
representative of the cases that are of general interest:
\begin{description}
   \item[\Case~1 :] There is no emission line in the spectrum.
   \item[\Case~2 :] There is a narrow Gaussian emission line at 2.85~keV
   with ${\rm SD}\footnote{SD stands for standard deviation.} = 0.04$~keV and
   equivalent width\footnote{The equivalent width is
   defined as ${\rm EW} = \lambda/[f(\theta^C,\mu)\cdot {\rm ARF} \cdot {\rm Exposure~time}]$.}, ${\rm EW} =
   0.198$~keV in the spectrum.
   \item[\Case~3 :] There is a moderate Gaussian emission line at 2.85~keV
   with ${\rm SD} = 0.21$~keV and ${\rm EW} = 0.661$~keV in the
   spectrum.
   \item[\Case~4 :] There is a narrow Gaussian emission line at 2.85~keV
   with ${\rm SD} = 0.04$~keV and ${\rm EW} = 0.659$~keV in the spectrum.
   \item[\Case~5 :] There are two narrow Gaussian emission lines
   with ${\rm SD} = 0.04$~keV, one at 1.20~keV with ${\rm EW} = 0.139$~keV
   and the other at 2.85~keV with ${\rm EW} = 0.198$~keV in the
   spectrum.
   \item[\Case~6 :] There are two narrow Gaussian emission lines
   with ${\rm SD} = 0.04$~keV, one at 1.20~keV with ${\rm EW} = 0.139$~keV
   and the other at 2.85~keV with ${\rm EW} = 0.659$~keV in the
   spectrum.
\end{description}
That is, each spectrum has either 0, 1, or 2 lines. Typical {\it
Chandra} data are recorded in an energy grid with bin width
0.01~keV. Thus, the narrow emission line (i.e., \Cases~2, 4, 5,
and 6) corresponds to about 17 energy bins and the moderate one
(i.e., \Case~3) about 85 energy bins, using $\pm2$ standard
deviations. As compared to the {\it Chandra} resolution, the delta
function emission line profile corresponds to 1 energy bin and
thus it does not correctly specify the width of the Gaussian
emission lines in this simulation. Through the simulation study,
however, we illustrate possible advantage of this model
misspecification in producing valid and efficient estimates and
associated uncertainties for the line location. We show that using
delta functions emission lines in the model is a useful strategy
even when the true line occupies multiple bins.

For each of the six spectra, we generate twenty test data sets
(120 data sets in total) each with about 1500 counts similar to
the observed number of counts in the {\it Chandra} X-ray spectrum
of PG1634+706 analyzed in \S\ref{park:sec:quasar}, mimicking the
real data situation. Each spectrum has a power law continuum with
$\alpha^C=3.728e$$-$5 and $\beta^C=1.8$. Our simulation is done
with Sherpa software \citep{free:etal:01} in CIAO\footnote{The
software is publicly available on {\tt
http://cxc.harvard.edu/ciao}.}, assuming the {\it Chandra}
responses (effective area and instrument response function) and no
background contamination.

\subsection{Validity of Delta Function Line Profiles}
\label{park:sec:simul1}

\begin{figure}[p]
\begin{center}
  \includegraphics[width=6in]{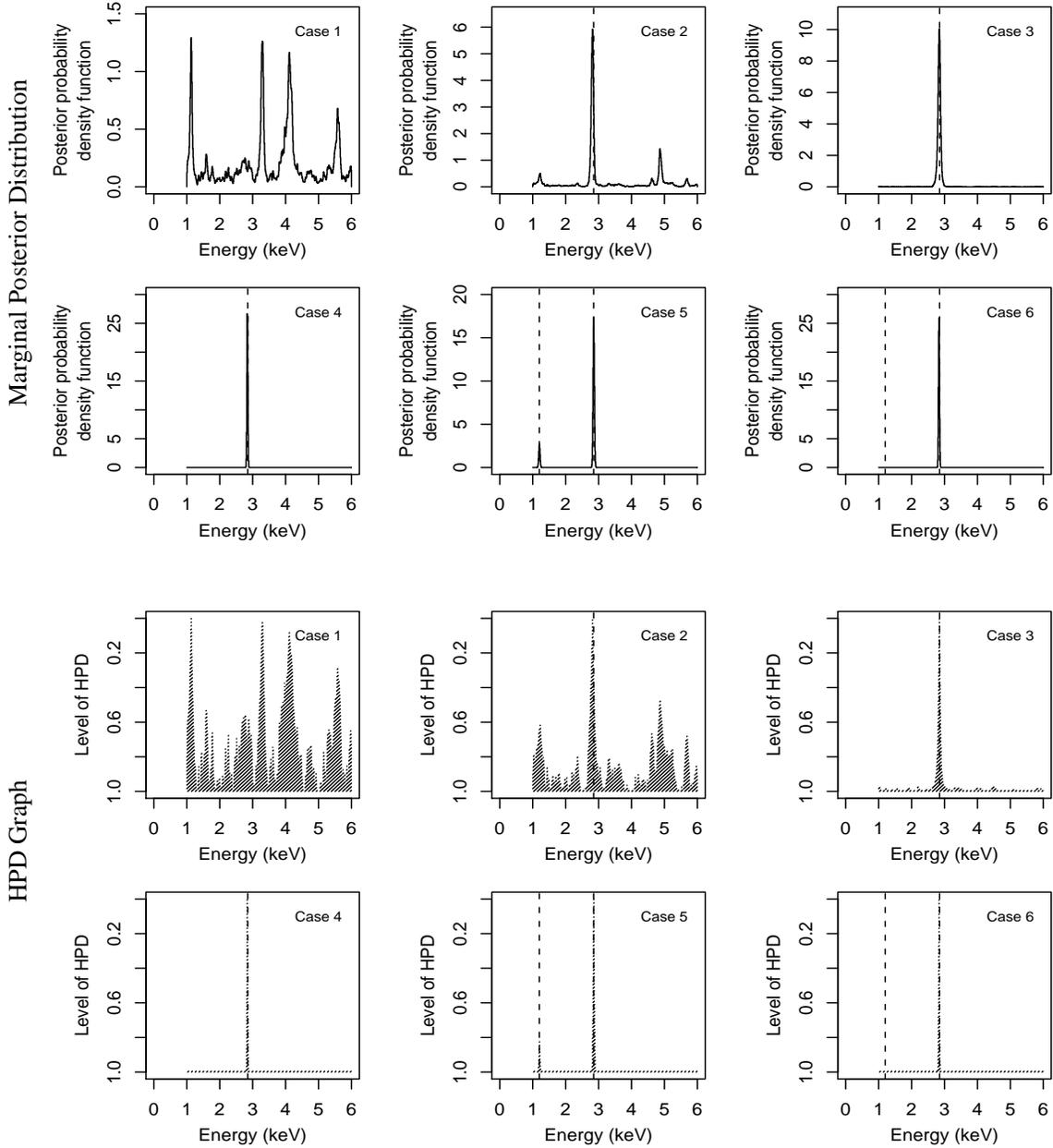}
    \caption{The Marginal Posterior Distribution and HPD Graph of
    the Line Location for One Simulation Under
    Each of the Six Cases in the Simulation
    Study. There is no emission line in the spectrum of \Case~1; the
    spectra of \Cases~2, 3, and 4 include one narrow or moderate
    emission line; and the spectra of \Cases~5 and 6 include two
    narrow emission lines. The top two rows illustrate the smoothed
    marginal posterior distributions of the line location for the six
    spectra in the simulation study, computed using Gaussian kernel
    smoothing with standard deviation 0.01~keV. The corresponding
    HPD graphs constructed with 100 HPD regions are presented in
    the bottom two rows; refer to Figure~\ref{park:fig:HPD-illust}.
    The vertical dashed lines represent the locations of the true emission
    lines.}
    \label{park:fig:simul}
\end{center}
\end{figure}

For each of the test data sets, we run state-of-the-art MCMC
samplers to fit a spectral model with a single delta function
emission line. Based on the Monte Carlo draws collected from the
multiple chains of the MCMC samplers, the top two rows of
Figure~\ref{park:fig:simul} present the marginal posterior
distribution of the delta function line location for one
simulation under each of the six cases; the vertical dashed lines
represent the true line locations. The marginal posterior density
is smoothed using Gaussian kernel smoothing with standard
deviation 0.01~keV, as described in \S\ref{park:sec:mm}.

When there is no emission line in the spectrum (i.e., \Case~1),
the posterior distribution of the delta function line location is
highly multimodal. In the case of a weak narrow Gaussian emission
(i.e., \Case~2), the marginal posterior distribution often remains
highly multimodal, but one mode typically identifies the true line
location. In practice, the local mode(s) of such a highly
multimodal posterior distribution may suggest plausible line
locations and show evidence for multiple lines; see
\S\ref{park:sec:simul2}. Even with a moderate line (i.e.,
\Case~3), the true line location appears well estimated with the
marginal posterior distribution of the delta function line
location. As we shall see in Table~\ref{park:tbl:simulHPD},
however, in this case the resulting posterior region under-covers
the true values because the true line is 85 times wider than the
specified model. With the strong narrow Gaussian line (i.e.,
\Case~4), the posterior distribution of the line location tends to
be unimodal, and the posterior mode correctly identifies the true
line location. The posterior distribution for \Case~5 in
Figure~\ref{park:fig:simul} is bimodal, with the modes
corresponding to the two true line locations. When multiple lines
are present in a spectrum, the posterior distribution of the
single line location can be multimodal, as shown in the \Case~5 of
Figure~\ref{park:fig:simul}. Thus the multiple modes may be
indicative of multiple lines; see \S\ref{park:sec:simul2} for
details. When one of two narrow Gaussian emission lines is much
stronger (i.e., \Case~6), the single delta function line model
tends to identify only one of the two true line locations. To
visualize the uncertainty of the fitted delta function line
location(s), the bottom two rows of Figure~\ref{park:fig:simul}
show the HPD graphs constructed with 100 HPD regions as described
in \S\ref{park:sec:mm}.

\subsection{Connection Between Multimodality and Multiple Lines}
\label{park:sec:simul2}

\begin{figure}[p]
\begin{center}
  \includegraphics[width=6in]{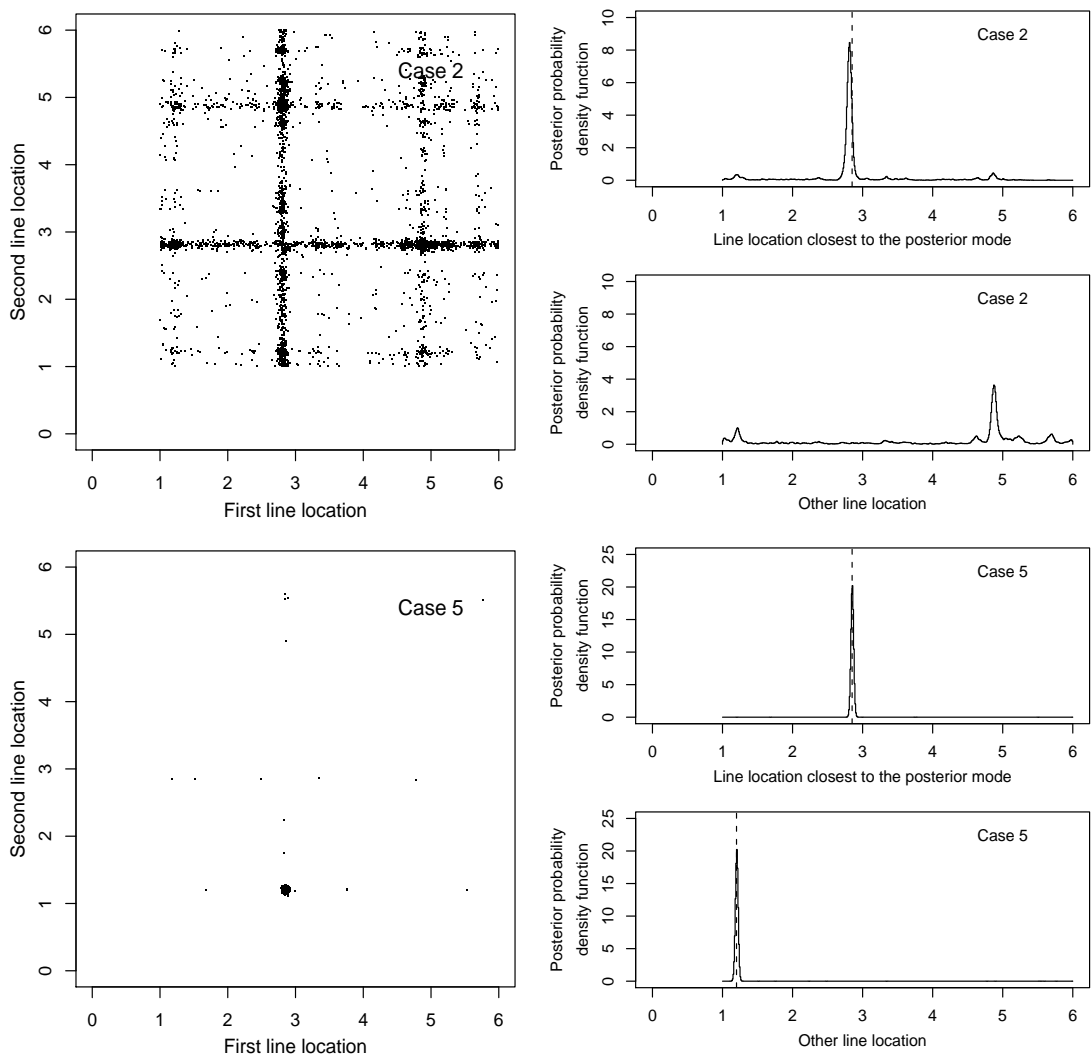}
    \caption{The Joint Posterior Distribution and the Corresponding
    Marginal Posterior Distributions of Two Line Locations in
    \Case~2 (one narrow emission line) and \Case~5 (two narrow
    emission lines). The vertical dashed lines in the right
    panels represent the locations of true emission lines.}
    \label{park:fig:simul-twolines}
\end{center}
\end{figure}

The multimodality in the marginal posterior distribution of a
single line location may indicate the existence of multiple lines
in a spectrum, provided the lines are well separated. When a model
is fitted with one emission line, modes in the likelihood function
of the line location correspond to ranges of energy with excess
emission relative to the continuum. Multiple modes in the
likelihood indicate that there are multiple ranges of energy with
such excess emission. The height of the mode is indicative of the
degree of excess. Thus, if there are several emission lines, we
might expect to see several corresponding modes in the likelihood.
If there is one energy range that dominates in terms of excess
emission, however, it corresponds to the dominate mode of the
likelihood. Thus, if there are lines of very different
intensities, only the strongest ones may show up as a mode of the
likelihood. This can be seen by comparing \Case~5 and \Case~6 in
Figure~\ref{park:fig:simul}.

If there is evidence for multiple lines in a spectrum or if we
suspect multiple lines a priori, we can fit a model with two or more
lines. We illustrate this using simulated data under \Case~2 (one
narrow line) and \Case~5 (two narrow lines). Beginning with \Case~2,
the actual spectrum has only one line, but we investigate what
happens when we fit two lines to this data.  A scatterplot of the
two fitted line locations identified when fitting two emission lines
to one of the data sets generated under \Case~2 is presented in the
top left panel of Figure~\ref{park:fig:simul-twolines}. There is a
label switching problem between the two fitted line locations
because of the symmetry of the emission lines in the model. We can
remove the symmetry by imposing a constraint on line locations. To
do this, we first fit the model with a single delta function line
profile and compute the posterior mode of its line location.
Returning to the model with two fitted delta functions, we separate
the two fitted line locations by setting the ``first" line location
to be the one closest to the posterior mode. In Case~2, the
posterior mode for the single line location is 2.815~keV, so that
the first line location is the line location closest to 2.815~keV
and the second line is the other location. The two panels in the top
right corner of Figure~\ref{park:fig:simul-twolines} show the
resulting marginal posterior distributions of the two fitted line
locations. As shown in the figure, the marginal posterior
distribution of the first line location correctly identifies the
true line location.  The spectrum used to generate the data under
\Case~2 has no second line, so that the marginal posterior
distribution of the second line is highly multimodal. In practice,
we may take the local mode(s) in the second marginal posterior
distribution as candidates for another line location. However, the
resulting HPD regions for the second line are wide, indicating that
either there is no second line or if there is, it cannot be well
identified.

When two emission lines are present in a spectrum, we follow the
same procedure as illustrated using the data generated under
\Case~5. The bottom left panel of
Figure~\ref{park:fig:simul-twolines} shows the scatterplot of two
line locations identified in the spectrum of \Case~5.
Label-switching is handled as above using the posterior mode for
the single line location, which is computed as 2.855~keV. The
fitted marginal posterior distributions of the first and second
line locations are given in the bottom right corner of
Figure~\ref{park:fig:simul-twolines}. When there are two emission
lines in a spectrum, the two true line locations are precisely
specified by the two marginal posterior distributions. This is a
verification of what is suggested by the multiple modes in the
marginal posterior distribution of the single line location shown
in Figure~\ref{park:fig:simul}.

\subsection{Possible Advantage of Model Misspecification}
\label{park:sec:simul3}

\begin{figure}[p]
\begin{center}
  \includegraphics[width=6in]{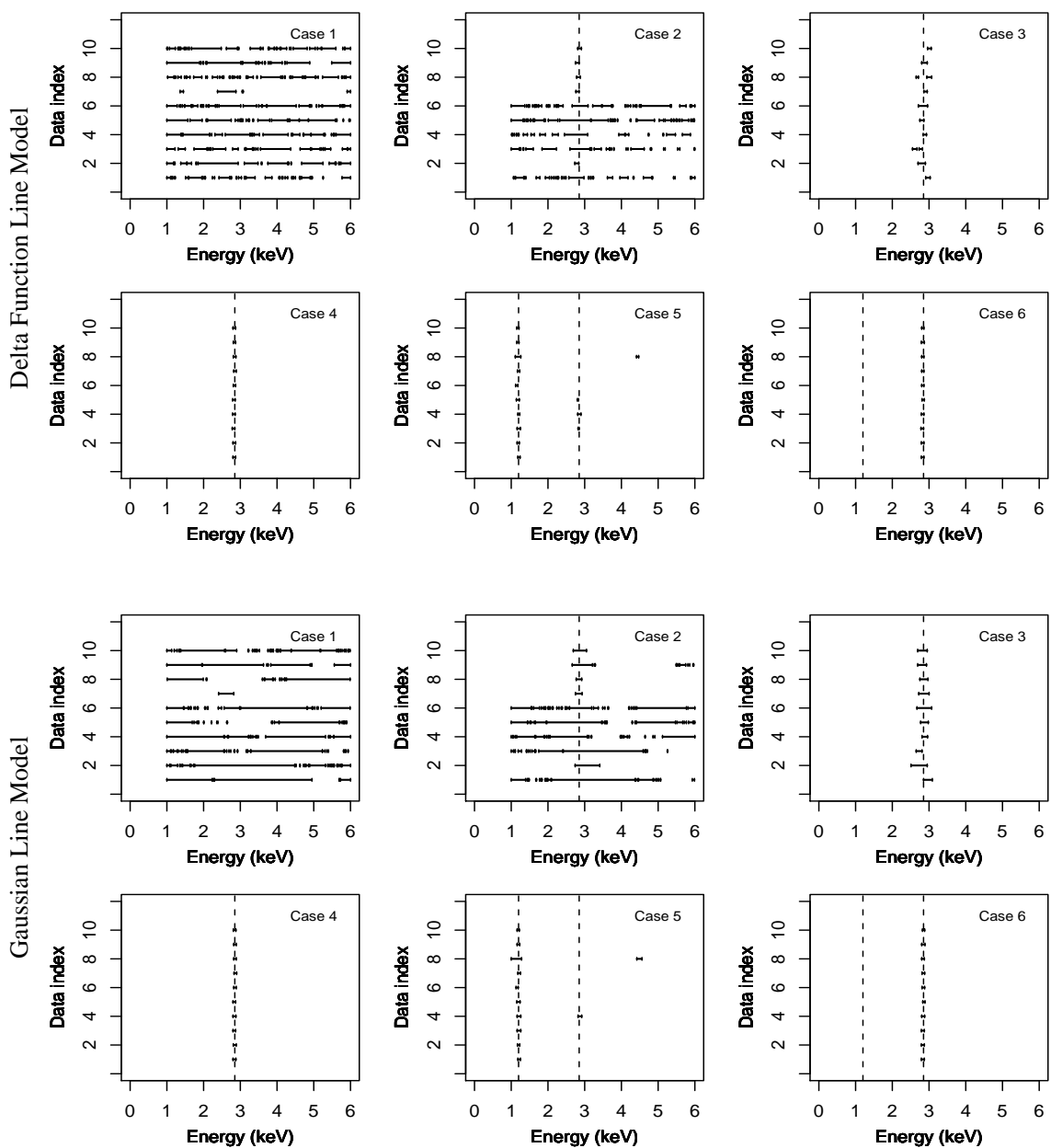}
    \caption{95\% HPD Regions for the Delta Function Line Location and
    Gaussian Line Location in the Simulation Study. The HPD regions
    are all computed under a model containing a single emission
    line and plotted against the index of the first ten
    simulated data sets. %Except in Case~1,
    The intervals produced when using a delta function line profile
    tend to be somewhat narrower and more precise. The vertical dashed
    lines represent the locations of true emission lines.}
    \label{park:fig:simul-10datasets}
\end{center}
\end{figure}

The possibility of model misspecification when using a delta
function to model an emission line depends on both the width of
the true line and the resolution of the detector. Misspecification
only occurs when the line is not contained in one energy bin, and
we shall illustrate that such misspecification only has
statistical consequences for the fitted line location if it is
very severe. Indeed there can be a possible statistical advantage
of using a delta function rather than a Gaussian line if we know
the spectral line is not too wide. As a toy example, consider a
simple Gaussian model with known standard deviation: When
$Y\sim{\rm N}(\mu,\sigma)$, a 95\% confidence interval for $\mu$
is given by $Y\pm1.96\sigma$. If we misspecify the modal as
$Y\sim{\rm N}(\mu,\varsigma)$ with $\varsigma<\sigma$, the
resulting interval for $\mu$ is shorter and has lower coverage. We
similarly underrepresent the error bars of an emission line
location when we use a delta function for a line that is not
contained in one energy bin. We expect this to reduce both the
length and coverage of the confidence regions. The advantage or
disadvantage of this strategy is not immediately clear, however,
since the nominal coverage of the intervals is based on an
asymptotic Gaussian approximation to the posterior distribution
which clearly does not apply in this setting. Nonetheless, our
simulation study illustrates that the use of a delta function line
profile can result in a shorter and more informative HPD region
while maintaining good coverage.

We now turn to the computation of HPD regions and the possible
statistical advantage of using delta functions in place of narrow
Gaussian emission lines. We fit a spectral model that includes a
single delta function line or a single narrow Gaussian line to the
twenty simulated data sets generated under each of the 6 cases.
After smoothing the marginal posterior distribution using Gaussian
kernel smoothing, we construct 95\% HPD regions for the line
location, as shown in Figure~\ref{park:fig:simul-10datasets}. For
visual clarity, we present results only for the first ten simulated
data sets in Figure~\ref{park:fig:simul-10datasets}; results from
all twenty simulated data sets are discussed in
Table~\ref{park:tbl:simulHPD}. Because there is no emission line in
the spectrum used to simulate data under \Case~1, the 95\% HPD
regions for the line location are very wide and show large
uncertainties for the fitted line location. When there is at least
one strong emission line (i.e., \Cases~3, 4, and 6), both line
models produce comparable HPD regions, although those computed under
the Gaussian line model appear somewhat wider. The tradeoff between
the two line models becomes evident when there is no strong emission
line in the spectrum (i.e., \Cases~2 and 5). In \Case~2, the 95\%
HPD regions for a single Gaussian line location are somewhat wider.
With the same nominal level, the delta function line model yields
more compact and informative HPD regions. An added advantage of the
delta function line model occurs in \Case~5 when the 95\% HPD
regions for a single delta function line location consist of two
disjoint HPD intervals which simultaneously contain the two true
line locations; this behavior is more often observed with the delta
function line model (2 times out of 20) than the Gaussian line model
(1 time out of 20).

\begin{table}[t]
  \begin{center}
  \caption{Summary of 95\% HPD Regions for the Line Location in the
  Simulation Study.}
  \medskip
  \begin{tabular}{ccccccc} \hline \hline
  & & \multicolumn{2}{c}{Delta function line} & & \multicolumn{2}{c}{Gaussian line}\\
  \cline{3-4}
  \cline{6-7}
  Case & Line type\footnotemark[5] & Coverage\footnotemark[6] & Mean length & & Coverage\footnotemark[6] & Mean length  \\
  \hline
   1 & no lines              & NA    & 3.354 & & NA    & 3.613\\
   2 & one narrow line       & 85\%  & 0.722 & & 100\% & 1.489\\
   3 & one moderate line     & 65\%  & 0.164 & & 95\%  & 0.263\\
   4 & one narrow line       & 100\% & 0.068 & & 100\% & 0.075\\
   5 & two narrow lines      & 95\%  & 0.081 & & 95\%  & 0.137\\
   6 & two narrow lines      & 100\% & 0.068 & & 100\% & 0.076\\
  \hline
  total & narrow line(s)     & 95\% & 0.235 & & 98.8\% & 0.444\\
  \hline\hline
    \label{park:tbl:simulHPD}
  \end{tabular}
  \end{center}
\end{table}
  \footnotetext[5]{The narrow emission lines are 17 bins wide
  (four standard deviations, i.e., 0.17~keV);
  moderate emission lines are 85 bins wide (i.e., 0.85~keV).}
  \footnotetext[6]{The coverage is the percentage of twenty 95\%
  HPD regions containing at least one true line location.}

To more closely inspect the advantage of model misspecification,
we evaluate the coverage of the true line locations and the mean
length of the 95\% HPD regions. For each of the 6 spectra,
Table~\ref{park:tbl:simulHPD} shows the percentage of the twenty
95\% HPD regions containing at least one true line location along
with the mean length of the regions. In \Case~1, there is no
emission line in the spectrum and there should therefore be great
uncertainties about the line location. Thus both line models
produce comparably wide HPD regions for the line location on
average. When a spectrum has at least one emission line, the delta
function line model yields HPD regions of smaller mean length and
with better coverage rates. The one exception is \Case~3 that
includes one moderate emission line whose location is
significantly undercovered with the delta function line model. In
this case, the 85-bin (i.e., 0.85~keV) wide line is very broad, as
compared to the delta function line profile which corresponds to
one energy bin (i.e., 0.01~keV). Although the delta function
appears to be efficient in identifying lines this wide, the model
underrepresents the uncertainty in their location. When all
emission lines are narrow (i.e., \Cases~2, 4, 5, and~6), the
misspecification of the line width seems to show an advantage. The
last row of Table~\ref{park:tbl:simulHPD} summarizes the 95\% HPD
regions when there is at least one narrow emission line in the
spectrum, i.e., an emission line with a width of 17 bins (i.e.,
0.17~keV). The delta function line model produces HPD regions with
53\% the mean length of HPD regions resulting from the Gaussian
line model, with coverage closer to the nominal 95\% rate.

Because fitting a spectral model involves MCMC sampling that is
computationally expensive and requires some supervision, exhaustive
simulations are difficult to carry out. In addition, the results may
depend on the line location, line strength, line width,
characteristics of the continuum, sample size, and so on. Based on
the simulation study, however, we conclude that the delta function
line model is useful for exploratory data analysis and for inference
when the true line is believed to be narrow. From a computational
point of view, an additional advantage is a significant reduction in
the computational time when using a delta function, owing largely to
the fact that the line width need not be fit. While the delta
function line model enjoys the advantages of model misspecification
when searching for the locations of narrow emission lines, it is not
designed to estimate the other line parameters. The delta function
emission line model underestimates the width, intensity, and EW of
the emission line. When making inference for these line parameters,
the delta function emission line model should not be used unless an
emission line is truly narrow relative to the resolution of the
detector.

\subsection{Summary of Line Parameters}
\label{park:sec:simul4}

\begin{figure}[p]
\begin{center}
  \includegraphics[width=6.5in]{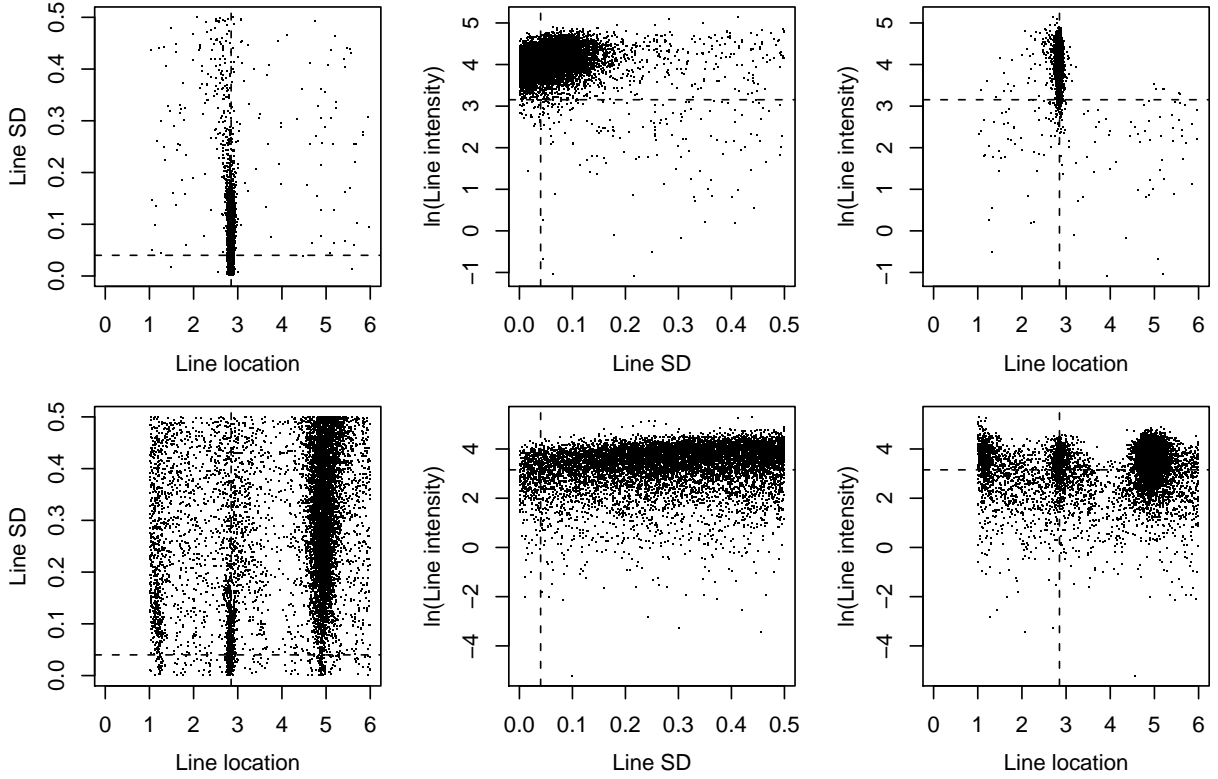}
    \caption{Joint Posterior Distributions of the Gaussian Line
    Location, the Line SD, and the Log of the Line Intensity
    from the Simulation Study. The two rows correspond to two
    typical types of test data in \Case~2, where the first row
    represents the test data with the line location having a unimodal
    posterior distribution and the second row the test data with the
    line location having a multimodal posterior distribution.
    The dashed lines represent either the true location (2.85~keV),
    the true standard deviation (0.04~keV), or the true log intensity
    (3.15 photons) of a narrow and weak Gaussian emission line
    in Case~2 of our simulation study.}
    \label{park:fig:simul-jntpost}
\end{center}
\end{figure}

When fitting a Gaussian emission line profile, the fitted line
intensity may be correlated with the fitted line width. To examine
the relationship among the Gaussian line parameters,
Figure~\ref{park:fig:simul-jntpost} shows the pairwise joint
posterior distributions of the Gaussian line location, the line
SD, and the log of the line intensity. The two rows in
Figure~\ref{park:fig:simul-jntpost} correspond to the results from
two simulated data sets, where the posterior distribution of the
line location tends to be unimodal or (highly) multimodal. The two
simulated data sets are both from \Case~2 but encompass the two
typical types of posterior distributions we see in \Cases~2, 3, 4,
5, and 6.

As the possible location of a Gaussian emission line is shifted
away from its posterior mode(s), the line width tends to increase
in order for the Gaussian emission line to be wide enough to
encompass the excess emission. When the line location is not well
specified, fitted values far from the true line location may be
consistent with the data so long as the width and equivalent width
of the line are sufficiently large. Thus, the joint posterior
distribution of the line location and line width tends to have
V-shape at each mode of the line location. This behavior is
illustrated in the left panels of
Figure~\ref{park:fig:simul-jntpost}. A larger fitted value of the
line width, in turn, increases the Gaussian line intensity,
resulting in an overall positive relationship in the middle panels
of Figure~\ref{park:fig:simul-jntpost}. As a result, the line
intensity tends to increase as the line location moves away from
each of its modes, as illustrated in the right panels of
Figure~\ref{park:fig:simul-jntpost}.

%%%%%%%%%%%%%%%%%%%%%%%%%%%%%%%%%%%%%%%%%%%%%%%%%%%%%% Updated AS, Dec.18

\section{Analysis of the Quasar PG1634+706}
\label{park:sec:quasar}

\subsection{The High Redshift Quasar PG1634+706}
\label{park:sec:quasar-desc}

X-ray spectra of many sources are available in {\it Chandra}
archives\footnote[7]{{\tt http://cxc.harvard.edu/cda.}}. We apply
our methods to the calibration source, PG1634+706 that was observed
6 times during the first year of the {\it Chandra} mission.  Only
one observation (\obs~1269) has been analyzed and published
\citep{haro:etal:07} and it indicates that a narrow iron line was
detected in this source. We include all available data sets in our
analysis to evaluate the location and significance of the line. We
focus on a single narrow emission line that might correspond to
either the narrow component or one of the two peaks of the broad
component, of the Fe-K-alpha line discussed in \S\ref{park:sec:sci}.
In principle an analysis might include two or three delta functions
in an effort to simultaneously identify all three features. As we
shall discuss, however, we find only marginal evidence for one
feature, and thus did not pursue the simultaneous fitting of
multiple features.

\begin{table}[t]
  \begin{center}
  \caption{Description of the {\it Chandra} Observations for PG1634+706}
  \medskip
  \begin{tabular}{ccc} \hline \hline
  Observed data set & Exposure time (sec.) & Total counts \\
  \hline
  \obs~47   & 5389.08 & 1651\\
  \obs~62   & 4854.57 & 1472\\
  \obs~69   & 4859.42 & 1457\\
  \obs~70   & 4859.68 & 1419\\
  \obs~71   & 4405.57 & 1356\\
  \obs~1269 & 10834.03& 2216\\
  \hline\hline
  \label{park:tbl:quasar}
  \end{tabular}
  \end{center}
\end{table}

PG1634+706 (redshift $z=1.334$) is a radio quiet and optically
bright quasar \citep{stei:sarg:91}. It is very luminous in X-rays
with the 2--10~keV band luminosity exceeding 10$^{46}$~erg~s$^{-1}$
\citep{jimen:etal:05}. The iron emission line in such luminous
sources is expected to be weaker than in lower luminosity AGN
\citep{nandra:etal:97}. The quasar was observed with ASCA
\citep{george:etal:00} and XMM-{\it Newton} \citep{page:etal:05} and
no line was detected at the energy of the 6.4~keV Fe-K-alpha line
(observed at ${\rm E}_{\rm obs}=2.738$~keV) with the limits on
equivalent width given as ${\rm EW}<750$~eV and ${\rm EW}<82$~eV,
respectively. However, the narrow line was detected in
\cite{haro:etal:07} analysis of one {\it Chandra} data at ${\rm
E}_{\rm obs}=2.84$~keV. Here we present all available {\it Chandra}
observations and search for the narrow emission line within the
available energy range.

PG1634+706 was observed with {\it Chandra} ACIS-S detector
\citep{weis:etal:02} as a calibration target six times on March 23
and 24, 2000. Each observation lasted between 4.4 and 11~ksec. We
use CIAO software\footnote[8]{{\tt http://cxc.harvard.edu/ciao.}} to
process the archival data and extracted the spectra assuming
circular source regions of 1.8~arcsec radius. We apply CALDB 3.3.0
calibration data.  Table~\ref{park:tbl:quasar} lists each
observation with its exposure time and the total counts in its
spectrum.

Below we apply our methods and search for the narrow emission line
in the available {\it Chandra} spectra.

\subsection{Fitting a Spectral Model}
\label{park:sec:fitting}

\begin{figure}[p]
\begin{center}
  \includegraphics[width=6in]{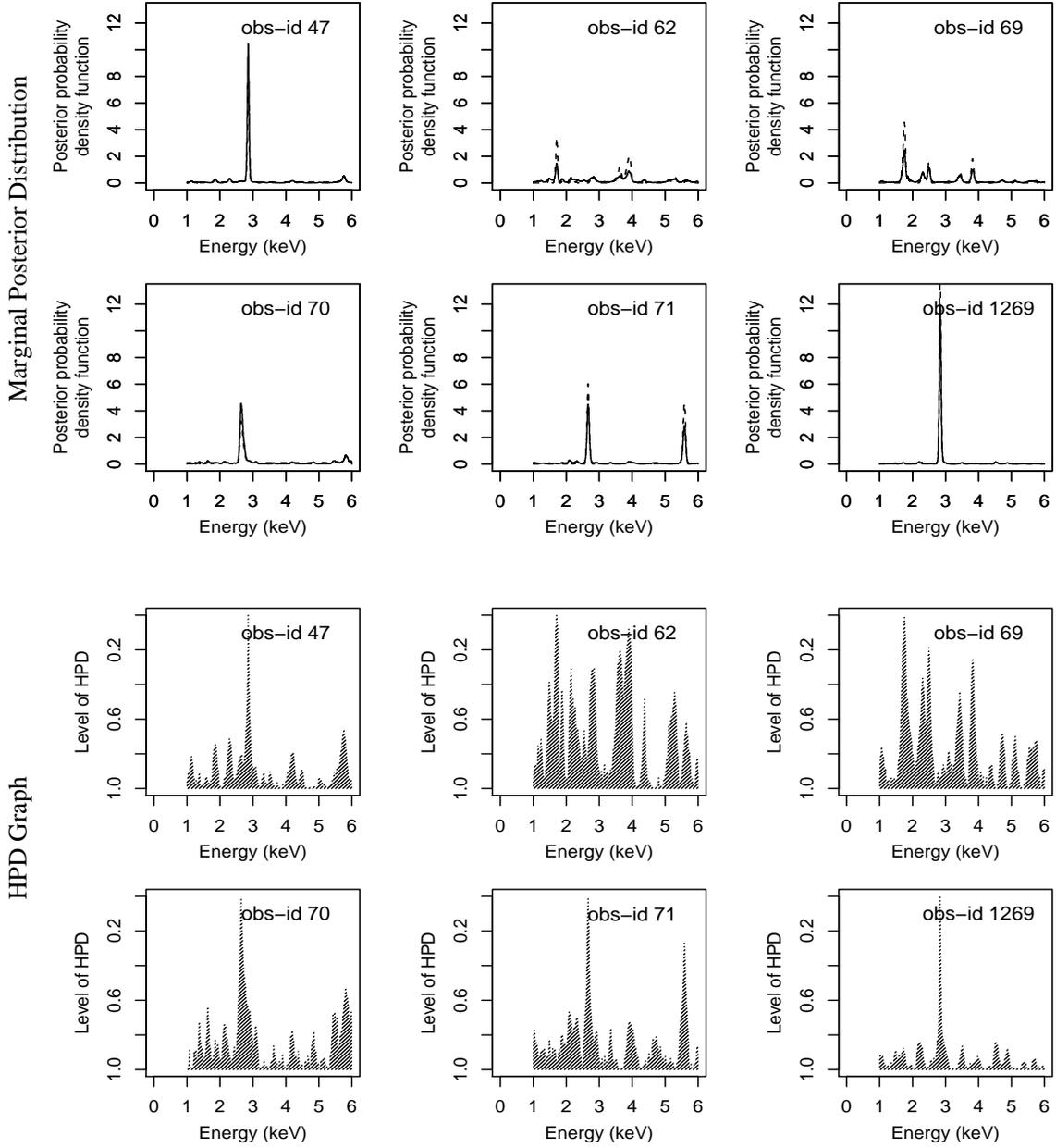}
    \caption{The Marginal Posterior Distribution and HPD Graph of the
    Delta Function Line Location, $\mu$, for Each of the Six
    Observations of PG1634+706. The solid lines in the first two rows
    represent the marginal posterior distribution of the delta
    function line location, and the dashed lines the profile posterior
    distribution. For several of the observations the two are
    nearly indistinguishable.}
    \label{park:fig:obs-quasar}
\end{center}
\end{figure}

We use our statistical algorithms using a delta function line
profile to search for a line in the {\it Chandra} spectra of
PG1634+706. When we look for emission lines, we typically confine
our attention to energies above 1~keV because we avoid regions
with potential calibration issues and effects related to
absorption.

The marginal posterior distribution of the line location for each
observation of PG1634+706 is computed using Monte Carlo draws and
is presented in the top two rows of
Figure~\ref{park:fig:obs-quasar}. The solid lines represent the
marginal posterior distributions, and the corresponding profile
posterior distributions are represented by dashed lines. Although
the marginal and profile posterior distributions differ in their
treatment of nuisance parameters, Figure~\ref{park:fig:obs-quasar}
illustrates that both representations capture similar peaks and
confirms that the Markov chain fully explores the parameter space
for the delta function line location. Because the marginal
posterior distribution of the delta function line location is
highly multimodal, we summarize the posterior distribution by
constructing an HPD graph to visualize the HPD regions of varying
levels; see \S\ref{park:sec:mm}. The HPD graph also illustrates
that the marginal posterior distribution is highly multimodal, so
that each HPD region may consists of a number of disjoint
intervals.
\begin{table}[t]
  \begin{center}
  \caption{95\% HPD Regions of the Delta Function Line Location.  The
  posterior modes of the line location near 2.74~keV where the
  Fe-K-alpha emission line is identified are indicated in bold face.}
  \medskip
  \begin{tabular}{cccc} \hline \hline
  Observed & Posterior & 95\% HPD & Posterior  \\[-.3em]
  data set & mode (keV) & region (keV) & probability\footnotemark[9]\\
  \hline
            & {\bf 2.885} & (2.44,\ 3.14) & 72.48\% \\
\up{\obs~47}& 5.915       & (5.44,\ 5.92) &  8.48\% \\
  \hline
            & 1.885       & (1.00,\ 1.97) & 25.88\% \\
            & {\bf 2.785} & (2.05,\ 3.02) & 21.65\% \\
\up{\obs~62}& 3.925       & (3.30,\ 4.06) & 28.27\% \\
            & 5.395       & (4.99,\ 5.41) &  8.17\% \\
  \hline
            & 1.955       & (1.51,\ 2.65) & 55.75\% \\
   {\obs~69}& 3.535       & (2.84,\ 3.62) & 12.41\% \\
            & 3.935       & (3.70,\ 4.01) & 11.79\% \\
  \hline
            & {\bf 2.795} & (2.37,\ 3.17) & 63.22\% \\
\up{\obs~70}& 5.945       & (5.34,\ 6.00) & 15.96\% \\
  \hline
            & 2.325       & (1.75,\ 2.45) &  8.81\% \\
   {\obs~71}& {\bf 2.815} & (2.50,\ 3.01) & 42.11\% \\
            & 5.625       & (5.38,\ 5.72) & 30.25\% \\
  \hline
   {\obs~1269}&{\bf 2.995}& (2.69,\ 3.08) & 84.96\% \\
  \hline\hline
  \label{park:tbl:HPDs}
  \end{tabular}
  \end{center}
\end{table}
\footnotetext[9]{Note that the posterior probability combined for
each \obs\ may not add up to 95\% because we list only posterior
intervals with posterior probability of 5\% or more.} %
For example, the 95\% HPD regions of the delta function line
location are presented in Table~\ref{park:tbl:HPDs} along with
local modes of the posterior distribution associated with each
interval. Each of the 95\% HPD regions is composed of a number of
disjoint intervals. Only the intervals that have posterior
probabilities greater than 5\% are presented in
Table~\ref{park:tbl:HPDs}, so that the probabilities may sum to
less than 95\%. For example, the two intervals of \obs~47
presented in Table~\ref{park:tbl:HPDs} have a combined posterior
probability of 80.96\% and the other eleven intervals not shown in
the table have a posterior probability of about 14.04\%, for a
total of 95\%. The posterior modes of the delta function line
location that are located near 2.74~keV are indicated in bold face
in Table~\ref{park:tbl:HPDs}.

\begin{figure}[t]
\begin{center}
  \includegraphics[width=6in]{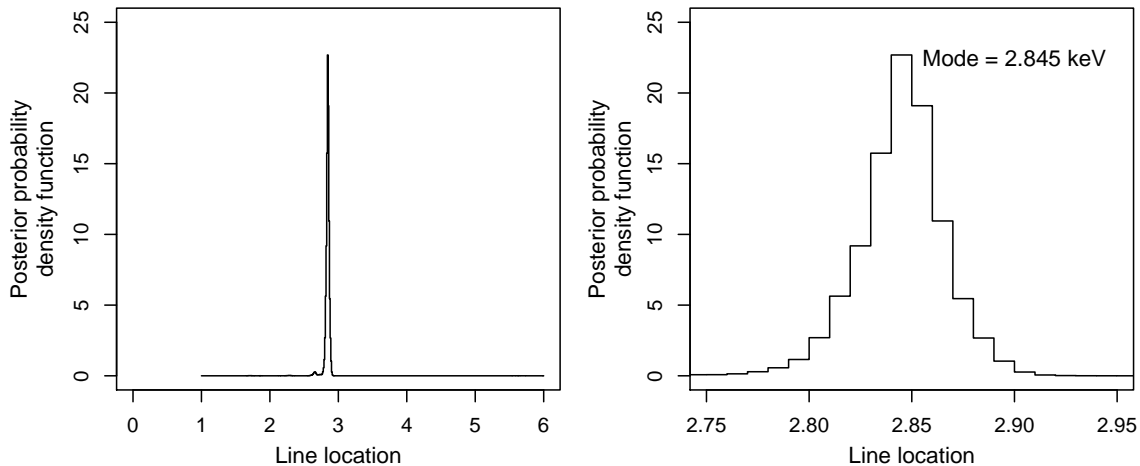}
    \caption{Posterior Distribution of the Line Location Given All
    of the Observations of PG1634+706. The left panel is
    plotted over the entire energy range of $\mu$ and the right panel
    over a range near 2.85~keV.}
    \label{park:fig:combined}
\end{center}
\end{figure}

The six observations of PG1634+706 were independently observed with
{\it Chandra}. Thus, under the flat prior distribution on $\mu$,
$p(\mu)\propto1$, the posterior distribution of the line location
given all six observations is given by
\begin{eqnarray}
  p(\mu|\mathbf{Y}^{\rm obs})
%   &=& \int p(\mu,\psi|y)d\psi \nonumber \\
    &\propto& \int\!\!\cdots\!\!\int \prod_{i=1}^6 L(\mu,\psi_i|\mathbf{Y}^{\rm obs}_i)
        d\psi_1 \cdots d\psi_6\nonumber \\
%   &=& \prod_{i=1}^6 \int L(\mu,\psi_i|Y_i)d\psi_i\nonumber \\
    &=& \prod_{i=1}^6 p(\mu|\mathbf{Y}^{\rm obs}_i),
  \label{park:eq:combined}
\end{eqnarray}
where $\mathbf{Y}^{\rm obs}=\{\mathbf{Y}^{\rm obs}_i,i=1,\dots,6\}$
denotes the six observations, $\mu$ denotes the delta function line
location parameter, $\psi=\{\psi_i,i=1,\dots,6\}$ denotes the set of
model parameters other than $\mu$ for the six observations, and
$L(\mu,\psi_i|\mathbf{Y}^{\rm obs})$ represents a likelihood
function of $(\mu,\psi_i)$ given $\mathbf{Y}^{\rm obs}$. (Here we
allow $\psi_i$ to vary among the six observations; i.e., we do not
exclude the possibility that the six observations have somewhat
different power law normalizations and photon indexes.) The values
of the posterior distribution given one of the individual data sets
is sometimes indistinguishable from zero because of numerical
inaccuracies. Thus we add 1/15000 to the posterior probability of
each energy bin and renormalize each of the posterior distributions.
This allows the product given in Equation~\ref{park:eq:combined} to
be computed for each energy bin and is somewhat conservative as it
increases the posterior uncertainty corresponding to each of the
individual data sets. Figure~\ref{park:fig:combined} presents the
marginal posterior distribution of the delta function line location
given all six observations computed in this way; the left panel
examines the whole range of the line location while the right panel
focuses on the range near 2.74~keV. As shown in
Figure~\ref{park:fig:combined}, the posterior distribution given all
six observations is fairly unimodal and symmetric, except the little
local mode near 2.655~keV. Thus, using a nominal 95\% HPD region,
the most probable delta function line location is summarized as
$2.845_{-0.055}^{+0.045}$~keV with posterior probability of 95.3\%.

\subsection{Model Checking and Evidence for the Emission Line}
\label{park:sec:model-check}

Posterior predictive methods \citep{rubi:81, rubi:84, meng:94a,
gelm:meng:96, gelm:meng:ster:96} can be employed to check the
specification of the spectral model.  This methods aim to check the
self-consistency of a model, i.e., the ability of the fitted model
to predict the data to which the model is fit. To evaluate and
quantify evidence for the inclusion of an emission line in the
spectrum, we extend the method of posterior predictive p-values
proposed by \citet{prot:etal:02} and \citet{vand:kang:04}.

With the {\it Chandra} observations of PG1634+706, we consider the
same spectral model discussed in \S\ref{park:sec:model} except that
we compare three models for the emission line:
\begin{description}
  \item[\model~0 :] There is no emission line in the spectrum.
  \item[\model~1 :] There is a delta function emission line with
  location fixed at 2.74~keV but unknown intensity in the spectrum.
  \item[\model~2 :] There is a delta function emission line with unknown location
  and intensity in the spectrum.
\end{description}
We could equally well consider a Gaussian line profile in \models~1
and 2; either line profile model results in a valid test. We
consider a delta function line profile simply because we are looking
for evidence of a narrow emission line.

\begin{figure}[t]
\begin{center}
  \includegraphics[width=6in,angle=0]{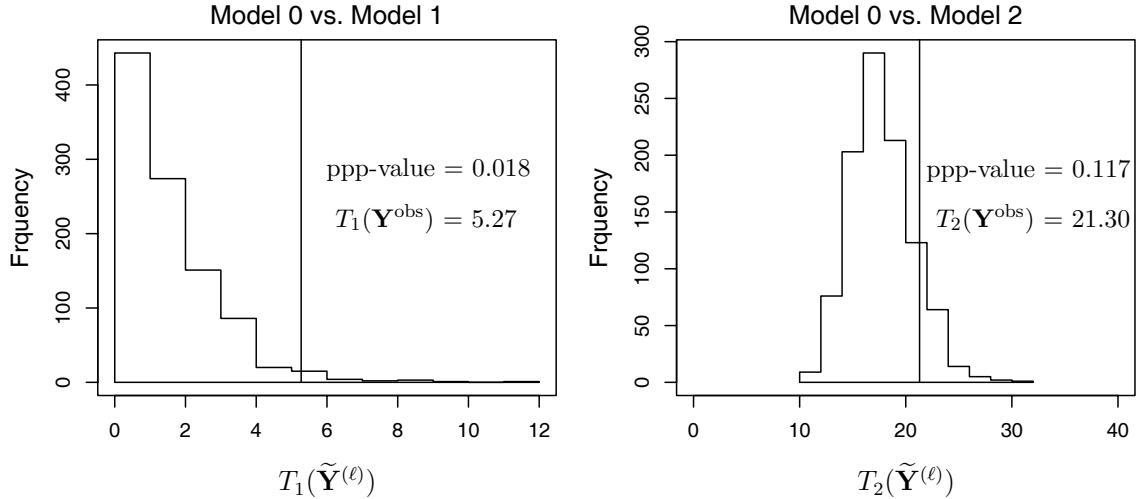}
    \caption{Posterior Predictive Checks Given All Six Observations
    of PG1634+706. In each of the two histograms, the observed test statistic
    (the vertical line) is compared with the test statistics
    from 1000 posterior predictive simulated data sets. The ppp-value
    is the proportion of the test statistics computed using the
    data simulated under \model~0 that are as extreme as or more
    extreme than the observed test statistic. Small ppp-values
    indicate stronger evidence for the emission line.}
    \label{park:fig:pp-check}
\end{center}
\end{figure}

We use ppp-values to compare the three models and quantify the
evidence in the data for the delta function emission line; see
\citet{prot:etal:02} for details of this method and its advantages
over the standard F-test, the standard Cash statistic
\citep{cash:79} (or likelihood ratio test statistic), and Bayes
Factors\footnote[10]{Because we are looking for evidence for the
emission line, we do not treat the two models under comparison
equally. That is, we are not simply looking for the model (with or
without the emission line) that best explains the data but we are
also attempting to guard against wrongly concluding that an emission
line is present when there is none. If we wished to simply compare
the two models, the deviance information criterion
\citep[DIC,][]{spie:etal:02} would be a better method to use.}. In
the posterior predictive checks, \model~1 fixes the delta function
line location at 2.74~keV using prior information as to the location
of the Fe-K-alpha emission line. In order to combine the evidence
for the line from all six observations with different exposure areas
and exposure times, we base our comparisons on the test statistic
that is the sum of the six loglikelihood ratio statistics for
comparing \model~$m$ and \model~0, i.e.,
\begin{eqnarray}
  T_m(\widetilde{\mathbf{Y}}^{(\ell)})=\sum_{i=1}^6\ln\Bigg\{
    \frac{\sup_{\theta\in\Theta_m}L\big(\theta|\widetilde{\mathbf{Y}}^{(\ell)}_i\big)}
    {\sup_{\theta\in\Theta_0}L\big(\theta|\widetilde{\mathbf{Y}}^{(\ell)}_i\big)}
    \Bigg\},\ \ m=1,\,2,\mbox{ and }\ell=1,\dots,1000,
\end{eqnarray}
where $\Theta_0$, $\Theta_1$, and $\Theta_2$ represent the
parameter spaces under \models~0, 1, and 2, respectively, and
$\widetilde{\mathbf{Y}}^{(\ell)}=\big\{\widetilde{\mathbf{Y}}^{(\ell)}_i,i=1,\dots,6\big\}$
denotes the collection of six data sets simulated under \model~0.
Specifically, we generate 1000 replications of each data set from
the posterior predictive distribution under \model~0 and compute
$T_m(\widetilde{\mathbf{Y}}^{(\ell)})$ for $m=1,2$. Histograms of
$T_1(\widetilde{\mathbf{Y}}^{(\ell)})$ and
$T_2(\widetilde{\mathbf{Y}}^{(\ell)})$ appear in
Figure~\ref{park:fig:pp-check}. Comparing the histogram of the
simulated test statistics with the observed value of the test
statistic yields the ppp-values \citep{rubi:84,meng:94a} shown in
Figure~\ref{park:fig:pp-check}. The ppp-value is the proportion of
the simulated test statistics that are as extreme as or more
extreme than the observed test statistic. Smaller ppp-values give
stronger evidence for the alternative model, i.e., \model~1 or
\model~2, thereby supporting the inclusion of the line in the
spectrum in our case. As shown in Figure~\ref{park:fig:pp-check},
there is evidence for the presence of the spectral line given all
six observations.  The comparison between \models~0 and~1 shows
stronger evidence for the line location because we are using extra
a priori information about the plausible line location.

%======================================================================

\section{Concluding Remarks}
\label{park:sec:conclusion}

This article presents methods to detect, identify, and locate narrow
emission lines in X-ray spectra via a highly structured multilevel
spectral model that includes a delta function line profile. Modeling
narrow emission lines with a delta function causes the EM algorithms
and MCMC samplers developed in \citet{vand:etal:01} to break down
and thus requires more sophisticated statistical methods and
algorithms. The marginal posterior distribution of the delta
function emission line location tends to be highly multimodal when
the emission line is weak or multiple emission lines are present in
the spectrum. Because basic summary statistics are not appropriate
to summarize such a multimodal distribution, we instead develop and
use HPD graphs along with a list of posterior modes. Testing for an
emission line in the spectrum is a notoriously challenging problem
because the value of the line intensity parameter is on the boundary
of the parameter space, i.e., zero, under a model that does not
include an emission line. Thus, we extend the posterior predictive
methods proposed by \citet{prot:etal:02} to test for the evidence of
a delta function emission line with unknown location in the
spectrum.

Using the simulation study in \S\ref{park:sec:simul}, we
demonstrate the potential advantage of model misspecification
using a delta function line profile in place of a Gaussian line
profile. We show that the delta function line profile may provide
more precise and meaningful summaries for line locations if the
true emission line is narrow. When multiple lines are present in
the spectrum, the marginal posterior distribution of a single
delta function line location may indicate multiple lines in the
spectrum.

Our methods are applied to the six different {\it Chandra}
observations of PG1634+706 in order to identify a narrow emission
line in the X-ray spectrum. Given all the six observations, the most
probable delta function line is identified at
$2.845_{-0.055}^{+0.045}$~keV in the observed frame. The
corresponding rest frame energy for the line is
$6.64_{-0.13}^{+0.11}$~keV, which may suggest the high ionization of
iron in the emitting plasma. There is some recent evidence that high
ionization iron line can be variable on short timescales \citep[see
for example Mkn766 in][]{miller:etal:06}. Such variability would
explain no detection of the emission line in one of the six {\it
Chandra} observations.

%BLoCXS (Bayesian fitting of Low Count X-ray Spectra) is free
%statistical software and will soon be available on the CIAO
%contributed software page.

%========================================================================
\acknowledgments

The authors gratefully acknowledge funding for this project partially
provided by NSF grant DMS-04-06085 and by NASA Contract NAS8-39073 and
NAS8-03060 (CXC). This work is a product of joint work with the
California-Harvard astrostatistics collaboration (CHASC) whose members
include J.~Chiang, A.~Connors, D.~van~Dyk, V.~L.~Kashyap, X.-L.~Meng,
J.~Scargle, A.~Siemiginowska, E.~Sourlas, T.~Park, A.~Young, Y.~Yu,
and A.~Zezas. The authors also thank the referee for many helpful and
constructive comments.

%========================================================================
\appendix

%========================================================================
\section{New Algorithms for Mode Finding and Posterior Simulation}
\label{ap:alg}

In this section we give an overview of the mode finding and
posterior simulation methods used to fit the spectral model with a
narrow emission line. Our summary is brief, and thus readers who
are interested in a more detailed description should refer to
\citet{vand:park:04,vand:park:08} and \citet{park:vand:08}.

%========================================================================
\subsection{Faster EM-type Algorithms for Mode Finding}
\label{ap:toyEM}

In order to illustrate our computational strategy, consider a
simplified example of an {\it ideal instrument} that is not subject to
the data contamination processes. In particular, the redistribution
matrix is an identity matrix, the effective area is constant, there
are no absorption features, and there is no background contamination.
In addition, we assume that the continuum is specified with no unknown
parameters and that there is a single Gaussian emission line that has
a known width $\nu_0$ and a known intensity $\lambda$. Thus, the line
location is the only unknown parameter, the source model given in
Equation~\ref{park:eq:ideal} simplifies to
\begin{eqnarray}
  \Lambda_j(\mu)=\Delta_jf(E_j)+\lambda \pi_j(\mu,\nu_0),
  \label{park:eq:simple}
\end{eqnarray}
and the counts are modeled as $Y_j\sim{\rm
Poisson}(\Lambda_j(\mu))$. This model can be fit using the method
of data augmentation \citep{tann:wong:87} by setting
$Y_j=Y_j^C+Y_j^L$, where $Y_j^C$ and $Y_j^L$ are the counts due to
the continuum and the emission line in bin $j$, respectively.  In
particular, the EM iteratively split the counts into continuum
counts and emission line counts. Given the current iterate of the
line location, $\mu^{(t)}$, the \Estep\ updates the line counts
via
\begin{description}
  \item[\Estep\ :] Compute ${\rm E}[Y_j^L|\:\mu^{(t)},\mathbf{Y}]$
    for each bin $j$. That is,
  \begin{eqnarray}
    \widehat{Y}_j^L\equiv
    {\rm E}\big[Y_j^L\big|\:\mu^{(t)},\mathbf{Y}\big] = Y_j
    \frac{{\lambda}\pi_j(\mu^{(t)},\nu_0)}
    {\Delta_j f(E_j)+{\lambda}\pi_j(\mu^{(t)},\nu_0)}.
  \label{park:eq:toy-Estep}
  \end{eqnarray}
\end{description}
Next, the \Mstep\ of EM updates the
emission line location $\mu^{(t+1)}$ by
\begin{description}
  \item[\Mstep\ :] Find $\mu^{(t+1)}={\sum_{j\in{\cal J}}
  E_j\widehat{Y}_j^L}/{\sum_{j\in{\cal J}}\widehat{Y}_j^L}$,
  \label{park:eq:toy-Mstep}
\end{description}
which is the weighted average of the bin energies and uses the
emission line counts as weights.  Although this EM algorithm is
simple, it breaks down when fitting the location of a {\it narrow}
emission line, i.e., when $\nu_0$ is small relative to the size of
the bins. In the extreme, the Gaussian line profile becomes a delta
function, so that $\pi_j(\mu^{(t)},\nu_0)$ is zero for all bins
except the bin containing $\mu^{(t)}$. This results in an E-step
that computes zero line counts in all bins except the bin containing
$\mu^{(t)}$ and finally an M-step that computes $\mu^{(t+1)} =
\mu^{(t)}$. This means that EM will return the same line location at
each iteration and that the algorithm will not converge to a mode.

In this simplified example, this difficulty can be avoided by directly
maximizing the posterior distribution
\begin{eqnarray}
  p(\mu|\mathbf{Y})
    \propto
    {\prod_{j\in{\cal J}}\{\Lambda_{j}(\mu)\}^{Y_j}e^{-\Lambda_{j}(\mu)}}.
  \label{park:eq:toy-obs-like}
\end{eqnarray}
Because of the binning of the data, we can treat possible line
locations within each bin as indistinguishable and compute the mode
by evaluating Equation~\ref{park:eq:toy-obs-like} on the fine grid
that corresponds to the binning of the data.

The situation is more complicated in the full spectral model
described in \S\ref{park:sec:model}. The method of data
augmentation can be used to construct efficient algorithms that
both fit the parameters in the continuum and the lines and account
for instrument response and background contamination. In the case of
the narrow emission line, however, we
must implement a strategy that uses less data augmentation when
updating the line location/width than when updating the other model
parameters.  The Expectation/Conditional Maximization Either
(ECME) algorithm \citep{liu:rubi:94} allows us to use no data
augmentation when updating the line location/width, but the resulting
M-step is time consuming owing to the multiple evaluations of the
conditional posterior distribution of the line location/width which
involve the large dimensional redistribution matrix, $\mathbf{M}$.  An
intermediate strategy uses the standard data augmentation scheme
to adjust for instrument response and background contamination but
does not separate continuum and line photons when updating the
line location/width. This strategy is an instance of the Alternating
Expectation/Conditional Maximization (AECM) algorithm
\citep{meng:vand:97} and each iteration is much quicker than with
ECME but more iterations are required for convergence. The
algorithms we use for mode finding aim to combine the advantages
of ECME and AECM by running one ECME iteration followed by $m$ AECM
iterations and repeating until convergence. \citet{vand:park:04}
call this a {\it Rotation($m$)} algorithm and illustrate the
computational advantage of the strategy.

%========================================================================
\subsection{Faster MCMC Samplers for Posterior Simulation}
\label{ap:toyMCMC}

Returning to the simplified example of Appendix~\ref{ap:toyEM}, we
can formulate a Gibbs sampler using the same data augmentation
scheme. Given the current iterate, $\mu^{(t)}$, \step~1 simulates
the line counts in bin $j$ via
\begin{description}
  \item[\step~1 :] Simulate $(Y_j^L)^{(t+1)}$ from
    $p(Y_j^L|\mu^{(t)},\mathbf{Y})$ for $j=1,\dots,J$. That is,
  \begin{eqnarray}
    Y_j^L|(\:\mu^{(t)},\mathbf{Y}) \sim
    {\rm Binomial}\Bigg(Y_j,\: \frac{{\lambda}\pi_j(\mu^{(t)},\nu_0)}
       {\Delta_j f(E_j)+{\lambda}\pi_j(\mu^{(t)},\nu_0)}\Bigg).
  \label{park:eq:step1}
  \end{eqnarray}
\end{description}
Next, \step~2 simulates the line location via
\begin{description}
  \item[\step~2 :] Simulate $\mu^{(t+1)}$ from
    $p(\mu|(\mathbf{Y}^L)^{(t+1)},\mathbf{Y})$. That is,
  \begin{eqnarray}
    \mu|\big((\mathbf{Y}^L)^{(t+1)},\mathbf{Y}\big) \sim
    {\rm N}\Bigg(\frac{\sum_{j\in{\cal J}} E_j(Y_j^L)^{(t+1)}}
            {\sum_{j\in{\cal J}}(Y_j^L)^{(t+1)}},\:
    \frac{\nu_0^2}{\sum_{j\in{\cal J}}(Y_j^L)^{(t+1)}}\Bigg).
  \label{park:eq:step2}
  \end{eqnarray}
\end{description}
We collect a posterior sample $\{\mu^{(t)},\: t=t_0+1,\dots,T\}$
after a sufficiently long burn-in period $t_0$; we discard the
burn-in draws, see \citet{vand:etal:01} for details.

When this algorithm is applied with a narrow emission line, it
breaks down just as the EM algorithm does. When a delta function is
used for the line profile, the simulation in
Equation~\ref{park:eq:step1} results in no line counts in any bin
except the one containing the line, and again $\mu$ does not move
from its starting value. When a narrow Gaussian line profile is
used, the situation is less extreme, but the sampler exhibits very
high autocorrelations and typically cannot jump among the posterior
modes. Just as with EM, the difficulty can be avoided by computing
the posterior distribution of $\mu$ on a find grid and directly
simulating
\begin{eqnarray}
  \mu^{(t)} \sim {\rm
  Multinomial}\Big\{1\:;\:\{p(\mu|\mathbf{Y})\big|_{\mu=E_j},j\in{\cal J}\}\Big\}.
  \label{park:eq:toy-DA}
\end{eqnarray}

When the method of data augmentation is used to account for data
contamination processes described in \S\ref{park:sec:model},
however, this approach should be modified in a manner analogous to
the ECME and AECM algorithms. This leads to the strategy of using
conditional distributions from different data augmentation schemes.
In this case, however, the resulting set of conditional
distributions used to construct the Gibbs sampler may be {\it
incompatible} and there may be no joint distribution that
corresponds to this set of conditional distributions. Although such
a sampler may result in efficient computation, care must be taken to
be sure the sampler delivers simulations from the target posterior
distribution. This is formalized through the Partially Collapsed
Gibbs (PCG) sampler of \citet{vand:park:08} which outlines the steps
that should be taken to ensure proper convergence; refer to
\citet{park:vand:08} for the applications and illustrations of PCG
samplers. The PCG samplers can be viewed as the stochastic version
of ECME and AECM, thereby allowing us to sample the line location
with no data augmentation (as in ECME) or partial data augmentation
(as in AECM). Thus, the PCG sampler differs from the Gibbs sampler
developed in \citet{vand:etal:01} in sampling the line location (and
line width), and in the order of sampling steps.
\citet{park:vand:08} design two PCG samplers to fit the spectral
model with a delta function emission line or a narrow Gaussian
emission line, which are called PCG~I and PCG~II, respectively. As
compared to PCG~I, PCG~II requires one additional sampling step for
the line width, so that fitting the narrow Gaussian emission line is
computationally more demanding than fitting the delta function
emission line.

\end{document}